\documentclass[twocolumn,amsmath,amssymb,floatfix,prd,nofootinbib]{revtex4-1}
\usepackage{epsfig}
\usepackage{dcolumn}
\usepackage{bbold} %
\usepackage{hyperref}
\usepackage{csquotes}

\usepackage[italic]{hepnames} %
\usepackage{hyperref}
\usepackage{tabularx}
\usepackage{siunitx} %

\DeclareMathOperator{\tQED}{\text{QED}}

\DeclareRobustCommand{\PQ}{\HepGenParticle{Q}{}{}\xspace} %
\DeclareRobustCommand{\PaQ}{\HepGenAntiParticle{Q}{}{}\xspace} %

\begin{document}
\title{Differential Heavy Quark Distributions and Correlations\\ in
Longitudinally Polarized Deep-Inelastic Scattering}
\author{Felix Hekhorn}
\email{felix.hekhorn@unimi.it}
\affiliation{Tif Lab, Dipartimento di Fisica, Universit\`a di Milano and\\ INFN, Sezione di Milano, Via Celoria 16, I-20133 Milano, Italy}
\author{Marco Stratmann}
\email{marco.stratmann@uni-tuebingen.de}
\affiliation{Institute for Theoretical Physics, University of T\"ubingen, Auf der Morgenstelle 
14, 72076 T\"ubingen, Germany}

\begin{abstract}
We present a first calculation of the heavy flavor contribution to the
longitudinally polarized deep-inelastic scattering structure function $g_1^{\PQ}$,
differential in the transverse momentum or the rapidity of the
observed heavy quark $\PQ$ or antiquark $\PaQ$.
All results are obtained at next-to-leading order accuracy in QCD within 
the framework of a newly developed parton-level Monte Carlo generator 
that also allows one to study observables associated with the produced heavy quark pair 
such as its invariant mass distribution or its correlation in azimuthal angle.
First phenomenological studies are carried out for various heavy quark 
distributions in a kinematic regime relevant for a future Electron-Ion Collider 
with a particular emphasis on the expected size of the
corresponding double-spin asymmetries and their sensitivity
to the still poorly constrained helicity gluon distribution.
Theoretical uncertainties associated with the choice of the factorization scale
are discussed for selected observables.
\end{abstract}

\maketitle

\section{Introduction and Motivation}
Heavy flavors, more precisely their contributions to deep-inelastic scattering (DIS) 
processes or their production in hadron-hadron collisions, have played a rather minor role so far
in constraining longitudinally polarized parton distributions (PDFs) \cite{Adolph:2012ca,Nocera:2013yia}
despite their sensitivity to the elusive gluon helicity density.
The prospect of a future high luminosity Electron Ion Collider (EIC) in the U.S.\ \cite{Boer:2011fh},
with its vastly extended kinematic reach compared with past and present fixed-target experiments 
\cite{Adolph:2012ca,Aidala:2012mv} will likely change this assessment considerably. 
For the first time, experiments will be able to probe the helicity structure of nucleons with 
unprecedented precision at momentum fractions $x$ well below the currently explored range $x\gtrsim 0.01$ 
and, at the same time, at energy scales $Q$ above $\SI{1}{\GeV}$, i.e., in the domain 
where perturbative QCD should be safely applicable and where contributions 
from heavy quarks (HQs) are potentially sizable.

Clearly, a proper treatment of HQs in polarized DIS will
become a must in any global QCD analysis of helicity PDFs 
\cite{deFlorian:2008mr,deFlorian:2014yva,Blumlein:2010rn} 
based on future EIC data \cite{Aschenauer:2012ve}.
Experiments at the DESY-HERA collider have shown in the past \cite{H1:2018flt} that the charm contribution 
to unpolarized DIS, i.e., to the structure function $F_2$, can reach about a level of $25\%$ at small $x$. 
Also, at not too large scales $Q$, as compared to the HQ mass $m$, one needs to
retain the full dependence on $m$ in all theoretical computations as a massless approximation,
i.e., treating charm quarks as massless partons in the proton, is not warranted here. 
This is primarily also the kinematic region of utmost significance for investigating polarized DIS
at the EIC \cite{Boer:2011fh}.

The presence of different scales, at least $m$ and $Q$, significantly complicates all
calculations involving HQs in DIS already at the level of next-to-leading order (NLO) 
accuracy, in particular, the evaluation of the necessary virtual corrections and phase space integrals.
This is even more so the case if one considers, in addition, the polarization of the initial-state
partons because of the well-known complications due to the presence of $\gamma_5$ and the
Levi-Civita tensor in the helicity projection operators in $n\neq 4$ dimensions \cite{Vogelsang:1990ug}.

An important step towards completing the already existing suite of NLO HQ production cross sections 
in longitudinally polarized hadro- \cite{Bojak:2001fx,Riedl:2009ye} and 
photoproduction \cite{Bojak:1998bd,Riedl:2012qc}
has been made recently in Ref.~\cite{Hekhorn:2018ywm} where the heavy flavor contributions
to the inclusive helicity DIS structure function $g_1$ have been computed
more than 35 years after the leading order (LO) expressions were first derived \cite{Watson:1981ce}.
This paper will build on these NLO results that were obtained with largely analytical methods
following closely the corresponding, pioneering calculations 
in the unpolarized case \cite{Smith:1991pw,Laenen:1992zk}. 

In the present work we shall develop a new parton-level Monte Carlo (MC) generator
that allows one to study systematically also more exclusive observables in longitudinally polarized DIS,
i.e., quantities which are differential in the transverse momentum $p_T$ or the rapidity $y$ 
of the observed heavy (anti)quark $\PQ$ ($\PaQ$) or which are associated with the 
produced HQ pair such as its invariant mass distribution or its correlation in azimuthal angle.
All phase space integrals are computed in four dimensions by means of 
standard MC techniques \cite{Kauer:2001sp}. Collinear and soft
divergencies at intermediate stages are made explicit by generalized plus distributions within the
framework of $n$-dimensional regularization \cite{tHooft:1972tcz} and cancelled upon adding 
the renormalized virtual corrections and appropriate factorization counter terms.
Contrary to our results in \cite{Hekhorn:2018ywm} and
thanks to the adopted subtraction method \cite{Ellis:1980wv}
there is no need to actually evaluate integrals in $n$ dimensions.  
 
We benefit greatly from previous applications of the subtraction method
to various HQ calculations at NLO accuracy
\cite{Mangano:1991jk,Frixione:1993dg,Harris:1995tu,Riedl:2009ye,Riedl:2012qc}, 
which provide essentially all of the formalism required here. In particular, 
we follow closely the methods and notations 
outlined in Refs.~\cite{Mangano:1991jk,Harris:1995tu} and shall only
very briefly review the main technical aspects in this paper in order 
to adopt them to the case of helicity-dependent DIS.  
We will utilize the results of our previous, largely analytical calculation \cite{Hekhorn:2018ywm} 
to validate our numerical results and the underlying MC code for fully inclusive 
and single-differential HQ processes 
which are amenable to both analytical and numerical methods.
We note that our code also provides unpolarized results. They appear in
the numerator of experimentally relevant double-spin asymmetries and, hence,
are needed in our phenomenological studies. Whenever possible, we use
existing unpolarized results in the literature 
\cite{Laenen:1992xs,Riemersma:1994hv,Harris:1995pr,Harris:1995tu}
for further validation of our MC code.  

Our MC implementation of polarized HQ electroproduction is in principle capable to produce 
any infrared safe observable, i.e., histograms for exclusive, semi-inclusive, or
fully inclusive quantities related to any of the outgoing particles or combinations
thereof. Experimental cuts can be implemented as well if needed, provided they can be
expressed in terms of the available partonic variables. The presented MC program and numerical
results are based on the virtual photon-hadron $(\Pggx h)$ center-of-mass system (c.m.s.)
frame, i.e., they make no reference to the lepton in DIS and are given solely in terms
of the relevant differential helicity structure function $dg_1$.  
In a next stage, our parton-level MC code can be straightforwardly 
expanded further, for instance, along the lines of the 
unpolarized \texttt{HVQDIS} code \cite{Harris:1997zq} by interfacing it with 
appropriate sets of HQ fragmentation functions to model the HQ decay into heavy mesons.

As a first phenomenological application of our MC code, we will explore in this paper
some HQ observables of potential interest for the physics program at a future EIC.
One of the main goals of the EIC \cite{Boer:2011fh} is to collect vital new information about 
the elusive helicity gluon density $\Delta \Pg$ from as many processes as possible 
and to exploit these data in future global QCD analyses. 
Since HQ electroproduction is driven by the LO photon-gluon fusion (PGF) process, 
$\HepProcess{\Pggx\Pg\to \PQ\PaQ}$, it is expected to play a major role in this exercise. 
In Ref.~\cite{Hekhorn:2018ywm} we have studied already the fully inclusive
contribution of charm quarks to the DIS structure function $g_1$ in the kinematic domain 
$x\lesssim 0.01$ and moderate $Q^2$ relevant for the EIC. Here, we will focus on various 
differential HQ distributions and correlations, including the experimentally relevant
double spin asymmetries, and compute them for the entire suite of helicity PDF 
uncertainty sets provided by the DSSV collaboration \cite{deFlorian:2014yva}.

It will turn out that the single inclusive distribution $dg_1/dp_T$ 
as well as the invariant mass spectrum of the produced HQ pair, 
$dg_1/dM$, are particularly sensitive to $\Delta \Pg$.
HQ correlations which are trivial at LO accuracy, for instance, the back-to-back
configuration in azimuthal angle, are also of phenomenological interest 
as they test higher order QCD corrections most clearly. These types of observables
are also particularly sensitive to all-order resummations, i.e., the treatment of
multiple soft gluon emission in the vicinity of, say, the back-to-back peak.
We should stress that these corrections are not included in our fixed order MC code and, hence,
results in these regions of phase space should be taken with a grain of salt;
soft gluon resummations in the context of HQ electroproduction have been reported in
Refs.~\cite{Laenen:1998kp}.
For the phenomenologically most interesting HQ observables we shall also investigate and estimate
the remaining theoretical uncertainties at NLO accuracy due to variations of the renormalization
and factorization scales.

Finally, we wish to remark that for phenomenological applications at a future EIC
at small $x$ and for $Q$ not much larger than the HQ mass $m$,
HQ production is most likely best described by retaining the full mass dependence
as has been done in our largely analytical calculation
\cite{Hekhorn:2018ywm} and for all numerical results presented here. 
This also implies that HQs can be only produced extrinsically, 
for instance, through the PGF mechanism and that the notion of a massless HQ PDF
makes no sense\footnote{For our purposes, we safely ignore here the 
possibility of a non-perturbative, ``intrinsic charm'' component in the
proton wave function \cite{Brodsky:1980pb}
which might play a role in the valence region at large momentum fractions $x$.}. 
However, in the asymptotic regime, $Q\gg m$, one might want to consider 
HQ parton densities which requires to set up some interpolating scheme that
matches a theory with $n_{lf}$ light flavors to a theory with $n_{lf}+1$ massless
flavors \cite{Buza:1996wv}. 
Different types of general-mass variable flavor number schemes (GM-VFNS) have been
proposed and adopted in global fits of unpolarized PDFs \cite{Butterworth:2015oua,Accardi:2016ndt}
but no such prescription has been considered so far for fits of helicity PDFs. 
Our results, along with the asymptotic limit $Q\gg m$ \cite{Buza:1996xr}, 
are necessary ingredients for a polarized GM-VFNS.

The remainder of the paper is organized as follows: in Sec.~\ref{sec:tech} we
briefly sketch the relevant technical aspects of setting up a parton-level
Monte Carlo generator for heavy quark production in longitudinally polarized DIS.
The phenomenological results are collected in Sec.~\ref{sec:pheno}. Here,
we first validate our MC code against our previous results based on largely
analytical methods. Next, we present detailed results for 
the transverse momentum and rapidity-differential single-inclusive distributions
of the longitudinally polarized DIS charm structure function $g_1^{\Pqc}$ 
and the experimentally relevant double-spin asymmetry $A_1^{\Pqc}$.
Finally, we turn to correlated HQ observables, most importantly, the
invariant mass distribution of the observed charm quark-antiquark pair
but also comprising their difference in azimuthal angle, their combined
transverse momentum, and the cone size variable.
The main results of the paper are summarized in Sec.~\ref{sec:summary}.

\section{Technical Aspects\label{sec:tech}}
In this Section, we will briefly review the main technical aspects
pertinent to our goal of setting up a parton-level MC program
for HQ production in polarized DIS. As was already mentioned,
we will heavily make use of existing methods adopted
for various other calculations of HQ production at NLO accuracy 
\cite{Mangano:1991jk,Frixione:1993dg,Harris:1995tu,Riedl:2009ye,Riedl:2012qc,Riedl:2014ywt}.
In particular, Refs.~\cite{Mangano:1991jk,Harris:1995tu} 
provide essentially all of the formalism required here, albeit for
the unpolarized case.

Photon-gluon fusion,
\begin{equation}
\HepProcess{\Pggx(q) \, \Pg(k_1) \to \PQ(p_1)\,\PaQ(p_2)}\,
\end{equation}
the sole mechanism for HQ electroproduction at LO accuracy \cite{Watson:1981ce},
receives both virtual and gluon bremsstrahlung $\mathcal{O}(\alpha_s)$ corrections,
\begin{equation}
\label{eq:pgf-nlo}
\HepProcess{\Pggx(q) \, \Pg(k_1) \to \Pg(k_2)\, \PQ(p_1)\,\PaQ(p_2)}\;,
\end{equation}
and is supplemented by a genuine NLO light [anti]quark-induced process
\begin{equation}
\label{eq:quark-nlo}
\HepProcess{\Pggx(q) \, \Pq(k_1) \,[\Paq(k_1)] \to \Pq(k_2) \,[\Paq(k_2)]\, \PQ(p_1)\,\PaQ(p_2)}\;,
\end{equation}
where the relevant four-momenta are labeled as $q$, $k_1$, $k_2$, $p_1$, and $p_1$. 
The photon is virtual, $q^2=-Q^2$, and the heavy quarks are taken to be massive,
$p_{1,2}^2=m^2$, while $k_{1,2}^2=0$. The spin-dependent matrix elements for all contributing 
$2\to2$ and $2\to3$ processes have been computed already in our previous
paper \cite{Hekhorn:2018ywm}, but rather than performing the needed phase space integrations
by largely analytical means we shall now utilize MC methods. This will enable us to
study any infrared safe observable in spin-dependent 
HQ electroproduction up to $\mathcal{O}(\alpha\alpha_s^2)$
even when experimental cuts are imposed. Here, $\alpha$ denotes the electromagnetic
and $\alpha_s$ the scale-dependent strong coupling. 
We shall stress already at this point that results obtained with
our parton-level MC code for HQ electroproduction will fail
whenever fixed-order results are bound to fail, i.e., when they
become sensitive to the emission of soft gluons; some examples
will be given in Sec.~\ref{sec:pheno}.

To compute all phase space integrals in four dimensions by means of 
standard MC techniques \cite{Kauer:2001sp} it is necessary
to cancel any intermediate soft or collinear singularities under the
integral sign beforehand. This is achieved by employing a variant of the subtraction method 
\cite{Ellis:1980wv,Mangano:1991jk,Harris:1995tu,Phaf:2001gc} and
expressing the matrix elements and the two- and three-particle phase space factors,
$d\mathrm{PS}_2$ and $d\mathrm{PS}_3$, respectively, in terms of the variables 
$x$, $y$, $\theta_1$, and $\theta_2$ that make the kinematic regions of soft and collinear 
emissions explicit.  
$x$ is the reduced invariant mass of the HQ pair, $s_5'\equiv s_5-q^2=(p_1+p_2)^2-q^2$
scaled by the reduced $\gamma^*$-parton c.m.s.\ energy squared $s'\equiv s-q^2=(q+k_1)^2-q^2$, 
i.e., $\rho^*\equiv (4m^2-q^2)/s'\le x\le 1$, and $-1\le y\le 1 $ is the cosine
of the angle between $\vec{q}$ and $\vec{k}_2$ in the frame where $\vec{q}+\vec{k}_1=0$.
Both $\theta_1$ and $\theta_2$ range between $0$ and $\pi$ and are used to
parametrize the spatial orientation of $k_{1,2}$ with respect to the plane
span by the other three momenta in the c.m.s.\ frame of the HQ pair.
They do not matter for the discussion of singular regions of phase space.

In the matrix element squared for (\ref{eq:pgf-nlo}) the emission of a soft
gluon corresponds to a $1/(1-x)$ singularity as $x\to 1$ and, likewise,
a collinear $1/(1+y)$ pole is encountered for $y\to -1$. 
The latter also appears in the diagrams contributing to process (\ref{eq:quark-nlo})
that are proportional to the electric charge of the HQ squared, $e_Q^2$.

The gist of the subtraction method adopted here \cite{Mangano:1991jk,Harris:1995tu} 
is to tame the singularities by multiplying the squared matrix elements 
in (\ref{eq:pgf-nlo}) and (\ref{eq:quark-nlo})
by an appropriate Lorentz-invariant factor that vanishes in the soft and/or collinear limit, yielding
finite functions that can be integrated numerically over $x$, $y$, $\theta_1$, and $\theta_2$.
The phase space $d\mathrm{PS}_3$, divided by the same factor, can be expressed in terms
of generalized plus distributions within the framework of $n=(4+\epsilon)$-dimensional regularization,
and all poles can be cancelled either upon adding the virtual contributions or by
applying a factorization counterterm before integration. We will briefly outline the most
important steps in what follows. For more details, we refer the reader to 
Refs.~\cite{Harris:1995tu,ref:phd}

Adopting our notation from Ref.~\cite{Hekhorn:2018ywm} for denoting the appropriate projections 
onto the different HQ contributions (i.e., structure functions) 
in unpolarized $(G,L)$ and longitudinally polarized $(P)$
DIS by a subscript $k=\{G,L,P\}$, 
we can define the modified $2\to 3$ matrix element squared 
for PGF (\ref{eq:pgf-nlo}) schematically as
\begin{equation}
\label{eq:mepgf}
{M'}_{k,\Pg}^{(1)} \equiv M_{k,\Pg}^{(1)} \left(\frac{{s'}^2}{2s}\right)^2\,(1-x)^2\,(1-y)\,(1+y)\;,
\end{equation}
where the the singularities in the limits $y\to -1$ and $x\to 1$ are now regulated.

To obtain the partonic cross section  $d\sigma^{(1)}_{k,\Pg}$ for the PGF process,
the matrix element squared in (\ref{eq:mepgf}) needs to be integrated over the 
appropriately rescaled phase space factor 
\begin{align}
\label{eq:ps3g}
d\mathrm{PS}_{3,\Pg} &\equiv d\mathrm{PS}_3 \,
   \left(\frac {2s}{{s'}^2}\right)^2\frac 1 {(1-x)^2(1-y)(1+y)} \nonumber\\
 &= \frac{T_\epsilon}{8 \pi^2}\, \frac{1}{\Gamma(1+\epsilon/2)} \left(\frac {{s'}^2} s\right)^{-1+\epsilon/2} 
    \left(\frac{s_5}{16\pi}\right)^{\epsilon/2} \nonumber \\
 &\times  \beta_5^{1+\epsilon}\,(1-x)^{-1+\epsilon}(1-y^2)^{-1+\epsilon/2}  \nonumber \\
 &\times  \sin^{1+\epsilon}(\theta_1)\, \sin^\epsilon(\theta_2)\, d\theta_1\, d\theta_2\, dy\, dx \,,
\end{align}
where
\begin{align}
\beta_5 &= \sqrt{1-\frac{4m^2}{s_5}}\;,\\
T_\epsilon &= \frac{1}{16\pi^2}\left[1 + \frac{\epsilon}{2}
\left[\gamma_E - \ln(4\pi)\right]\right] + \mathcal{O}(\epsilon^2)\;.
\end{align}
Combined with the usual flux and color-related prefactors and
\begin{align}
E_{k=G,L}(\epsilon)=\frac{1}{1+\epsilon/2},\; E_{k=P}(\epsilon)=1
\end{align}
this yields
\begin{align}
d\sigma^{(1)}_{k,\Pg} &= \frac{1}{2s'} \frac{E_k(\epsilon)}{16} {M'}_{k,\Pg}^{(1)}\,d\mathrm{PS}_{3,\Pg} 
\end{align}
which can be split up into three contributions \cite{Harris:1995tu}
\begin{align}
\label{eq:xsec-bare}
d\sigma^{(1)}_{k,\Pg} &= d\sigma^{(1),s}_{k,\Pg} +  d\sigma^{(1),c-}_{k,\Pg} +  d\sigma^{(1),f}_{k,\Pg}\;,
\end{align}
corresponding to the soft (s), collinear (c-), and finite (f) parts of $d\sigma^{(1)}_{k,\Pg}$, respectively.
This separation is based on expanding the expressions $(1-x)^{-1+\epsilon}$ and $(1-y^2)^{-1+\epsilon/2}$ 
in Eq.~(\ref{eq:ps3g}) in terms of generalized 
plus distributions \cite{Harris:1995tu,Mangano:1991jk,Frixione:1993dg}, for instance,
\begin{eqnarray}
(1-x)^{-1+\epsilon} &=& \left(\frac 1 {1-x}\right)_{\tilde\rho} 
    + \epsilon \left(\frac{\ln(1-x)}{1-x}\right)_{\tilde \rho} + \delta(1-x) \nonumber \\
 &\times& \left[\frac 1 \epsilon + 2\ln(\tilde\beta) 
    + 2\epsilon\ln^2(\tilde\beta)\right] + \mathcal{O}(\epsilon^2)\;,
\end{eqnarray}
where $\tilde{\beta}=\sqrt{1-\tilde{\rho}}$.
As usual, all $\tilde{\rho}$ and $\omega$ distributions are understood as an integral over a 
sufficiently smooth function $f(x)$ such that, e.g.,
\begin{align}
\int\limits_{\tilde\rho}^1\! dx\,f(x)\left(\frac 1 {1-x}\right)_{\tilde\rho} 
&= \int\limits_{\tilde\rho}^1\! dx\,\frac {f(x) - f(1)} {1-x}
\end{align}
holds.
The two regularization parameters can be taken anywhere in the range
$\rho^*\leq\tilde\rho < 1$ and $0<\omega\leq 2$. The specific choice 
only affects the rate of convergence and stability of the numerical MC integrations.
For all our purposes we can adopt the same values as in the corresponding 
unpolarized \texttt{HVQDIS} code \cite{Harris:1995tu,Harris:1995pr,Harris:1997zq}.

$d\sigma^{(1),s}_{k,\Pg}$ and $d\sigma^{(1),c-}_{k,\Pg}$ exhibit soft and collinear
singularities, respectively, that have been made manifest in $n$ dimensions by
introducing the $\tilde{\rho}$ and $\omega$ prescriptions. 
To proceed, one notices that the $1/\epsilon$ pole in $d\sigma^{(1),c-}_{k,\Pg}$
in the limit $y\to -1$ assumes the form dictated by the factorization theorem and, 
hence, can be absorbed
into the definition of the PDFs. To this end, one adds an appropriate
``counter cross section'' to $d\sigma^{(1),c-}_{k,\Pg}$ which is a convolution of the
$n$-dimensional LO gluon-gluon splitting function $P_{k,\Pg\Pg}^{(0),\tilde{\rho}}(x)$
and the Born cross section $d\sigma^{(0)}_{k,\Pg}$ evaluated at a shifted 
kinematics $x\, k_1$. The kernel $P_{k,\Pg\Pg}^{(0),\tilde{\rho}}(x)$ needs
to be expressed in terms of the generalized $\tilde{\rho}$ distribution and
reads in the polarized case $k=P$ \cite{Vogelsang:1995vh}
\begin{align}
P_{P,\Pg\Pg}^{(0),\tilde\rho}(x)
 &= 2C_A\left[\left(\frac{1}{1-x}\right)_{\tilde\rho} - 2x + 1 + \epsilon(x-1) \right]\nonumber\\
 &+ \delta(1-x)\left(\frac{\beta_0^{lf}}{2} + 4C_A\ln(\tilde\beta) - \epsilon\frac{C_A}{6}\right)
\end{align}
where $\beta_0^{lf}=\frac{11}{6}C_A-\frac{1}{3}n_{lf}$ with $C_A=3$ and $n_{lf}$ denoting the number
of active light quark flavors. The splitting function $P_{k,\Pg\Pg}^{(0),\tilde{\rho}}(x)$
can be decomposed into a soft (S) part, $\propto \delta(1-x)$, and a hard (H) part as well as into
four-dimensional and $\mathcal{O}(\epsilon)$ pieces as follows
\begin{align}
P_{k,\Pg\Pg}^{(0),\tilde{\rho}}(x) &= P_{k,\Pg\Pg}^{H,\tilde\rho,4}(x) 
+ \epsilon\; P_{k,\Pg\Pg}^{H,\tilde\rho,\epsilon}(x)  \nonumber\\
&+ \delta(1-x)\left[
P_{k,\Pg\Pg}^{S,\tilde\rho,4}(x) 
+ \epsilon\; P_{k,\Pg\Pg}^{S,\tilde\rho,\epsilon}(x) \right]
+ \mathcal{O}(\epsilon^2)\;. 
\label{eq:pgg}
\end{align}

Combining everything, the finite, factorized collinear 
cross section $d\hat{\sigma}_{k,\Pg}^{(1),c-}$
in the case $k=P$ reads
\begin{align}
d\hat{\sigma}_{P,\Pg}^{(1),c-} &= \alpha_s^2\,\alpha\, e_Q^2 
\frac{2\,\pi}{x\,s'} \,C_A\,C_F \,B_{P,\tQED}(x k_1)\,d\mathrm{PS}_2^{x} \nonumber\\
&\Bigg[(1-x) \,P_{P,\Pg\Pg}^{H,\tilde\rho,4}(x)
\Bigg\{\frac{1}{(1-x)_{\tilde\rho}}
\Bigg[ \ln \left(\frac{s'}{\mu_F^2}\right) \nonumber\\
&+ \ln\left(\frac{s'}{s}\right) + \ln \left(\frac{\omega}{2}\right)
\Bigg] \nonumber\\
&+ 2\left(\frac{\ln(1-x)}{1-x}\right)_{\tilde\rho} \Bigg\} 
+ 2 P_{P,\Pg\Pg}^{H,\tilde\rho,\epsilon}(x)\Bigg] 
\label{eq:sigmacoll}
\end{align}
where $d\mathrm{PS}_2^{x}\equiv d\mathrm{PS}_2|_{s\to s_5}\, dx$ 
denotes the relevant two-particle phase space factor for
collinear kinematics with
\begin{align}
d\mathrm{PS}_2(s) &= \frac{\beta\,\sin\theta_1}{16\pi\,\Gamma(1+\epsilon/2)} 
\left(\frac{s\,\beta^2\sin^2(\theta_1)}{16\pi}\right)^{\epsilon/2} d\theta_1,
\end{align}
$\beta=\sqrt{1-\rho}$, $\rho=4m^2/s$,
and $B_{P,\tQED}$ is  related to the Born matrix element squared for the PGF process, 
see, e.g., Eq.~(13) in Ref.~\cite{Hekhorn:2018ywm}. 
Equation (\ref{eq:sigmacoll}) carries a dependence on the arbitrary 
factorization scale $\mu_F$ where the collinear subtraction is performed.
We note, that for finite $Q^2$ there is no collinear divergence associated with
the limit $y\to 1$ that requires extra attention in the case of photoproduction.

Next, all remaining $1/\epsilon^2$ and $1/\epsilon$ singularities cancel upon 
combining the soft contribution $d\sigma^{(1),s}_{k,\Pg}$ with the 
virtual (loop) corrections $d\sigma^{(1),v}_{k,\Pg}$, renormalized at a scale $\mu_R$,
see Refs.~\cite{Hekhorn:2018ywm,ref:phd,Laenen:1992zk},
and the soft-collinear piece from the factorization counter term $\propto \delta(1-x)$
in (\ref{eq:pgg}).
For $k=P$ the finite soft plus virtual cross section reads
\begin{align}
d\hat{\sigma}^{(1),s+v}_{P,\Pg} &= \alpha_s^2\,\alpha\, e_Q^2 \frac{4}{s'}\,C_A\,C_F\,B_{P,\tQED} \Bigg[4C_A\ln^2(\tilde\beta) +\ln(\tilde\beta) \nonumber\\
&\times \Bigg\{ 
 2\, C_A \left[ \ln\left(\frac{-t_1}{m^2}\right)
 +\ln\left(\frac{-u_1}{m^2}\right)-\ln\left(\frac{\mu_F^2}{m^2}\right)\right] \nonumber\\
 &- 2\,C_F + \frac{s-2m^2}{s\,\beta}\ln(\chi)(C_A-2\,C_F) \Bigg\} + \frac {\beta_0^{lf}}{4}\nonumber\\
 &\times \left[ \ln\left(\frac{\mu_R^2}{m^2}\right)
  -\ln\left(\frac{\mu_F^2}{m^2}\right)\right] 
 + f_{P}(s,\theta_1) \Bigg] d\mathrm{PS}_2\;.
\end{align}
The result has no dependence on $x$ and $y$, nor on $\theta_2$.
The function $f_{P}$ contains all remaining finite contributions, mainly expressed 
in terms of various logarithms and dilogarithms, but independent of 
$\tilde\rho$, the scales $\mu_F$ and $\mu_R$, and $\beta_0^{lf}$.

Finally, the cross section $d\sigma^{(1),f}_{k,\Pg}$ in Eq.~(\ref{eq:xsec-bare})
collects all contributions that are finite in the limits $x\to 1$ and $y\to -1$.
It can be expressed in terms of the generalized distributions as follows
\begin{align}
d\sigma^{(1),f}_{P,\Pg} &= \left(\frac 1 {4\pi}\right)^4 \frac{1}{16s'} \frac{s\beta_5}{s'^2}
\left(\frac 1 {1-x}\right)_{\tilde\rho} \left(\frac 1 {1+y}\right)_\omega \frac 1 {1-y}
\nonumber\\
&\times {M'}_{P,\Pg}^{(1)}\,dx\,dy\sin(\theta_1)\,d\theta_1\,d\theta_2
\end{align}
where we have limited ourselves to $k=P$ as before.

In summary, the PGF process up to NLO accuracy can be expressed as
\begin{equation}
d\hat{\sigma}_{k,\Pg} = d\hat{\sigma}_{k,\Pg}^{(0)}+d\hat{\sigma}_{k,\Pg}^{(1),c-}+
d\hat{\sigma}^{(1),s+v}_{k,\Pg}+d\sigma^{(1),f}_{k,\Pg}
\label{eq:xsec-final}
\end{equation}
where now all terms are finite and, hence, all phase space integrations can be 
performed with MC methods in four dimensions. We note that each of the
three NLO contributions in (\ref{eq:xsec-final}) individually 
depends on $\tilde{\rho}$ and $\omega$ but not their sum.
To generate the predictions for HQ structure functions in DIS at the hadronic level, 
which will be shown and discussed in the next section, one needs to convolute 
the partonic result in (\ref{eq:xsec-final}) with the appropriate PDF,
i.e., in case of $k=P$ with a set of helicity-dependent PDFs. 

Finally, the genuine NLO contribution from initial-state light quarks
(\ref{eq:quark-nlo}) is conceptually much simpler to obtain than for
the PGF process. To this end, one notices that (\ref{eq:quark-nlo})
exhibits only a collinear singularity for $y\to-1$ and, hence,
one starts by defining the modified matrix element 
and appropriately rescaled phase space factor as
\begin{equation}
{M'}_{k,\Pq}^{(1)} \equiv M_{k,\Pq}^{(1)} \, \frac{{s'}^2}{2s}\,(1-x)\,(1+y)\;,
\end{equation}
and
\begin{align}
\label{eq:ps3q}
d\mathrm{PS}_{3,\Pq} &\equiv d\mathrm{PS}_3 \,
   \frac {2s}{{s'}^2} \frac{1}{(1-x)(1+y)} \,,
\end{align}
respectively.
As before, the divergent factor $(1+y)^{-1+\epsilon}$ that emerges
in $d\mathrm{PS}_{3,\Pq}$ can be regularized by introducing
a generalized plus distribution.

The collinear singularity in $d\sigma^{(1),c-}_{k,q}$
is lifted by adding a counter term containing
the LO quark-gluon splitting function $P_{k,\Pq\Pg}^{(0),\tilde{\rho}}(x)$
and the $d\sigma^{(0)}_{k,\Pg}(x\,k_1)$.
As a result, the finite light quark induced process at NLO accuracy can be 
written as
\begin{equation}
d\hat{\sigma}_{k,\Pq} = d\hat{\sigma}_{k,\Pq}^{(1),c-}+d\sigma^{(1),f}_{k,\Pq}
\end{equation}
where (for $k=P$)
\begin{align}
d\hat{\sigma}_{P,\Pq}^{(1),c-} &= \alpha_s^2\,\alpha\,e_Q^2 \,\frac{8\pi}{xs'} \, 
B_{P,\tQED}(x\,k_1)\, d\mathrm{PS}_2^{x}\,C_F \Big[(2-x)
\nonumber\\
&\times \left\{ \ln\left(\frac{s'}{\mu_F^2}\right) + \ln\left(\frac{s'}{s}\right) 
+ \ln\left(\frac{\omega}{2}\right) + 2\ln(1-x)\right\}\nonumber\\
&+ 2(x-1)\Big]
\end{align}
and
\begin{align}
d\sigma^{(1),f}_{P,\Pq}&= - \left(\frac 1 {4\pi}\right)^4 \frac{\beta_5}{12s'} 
 \left(\frac 1 {1+y}\right)_\omega \,{M'}_{P,q}^{(1)}\nonumber\\
&\times dx\,dy\,\sin(\theta_1)\,d\theta_1\,d\theta_2\;.
\end{align}
To obtain the contribution of (\ref{eq:quark-nlo}) to the
hadronic HQ structure functions one needs to perform the 
convolution of $d\hat{\sigma}_{k,\Pq}$ with appropriate combinations
of light quark PDFs.
\section{Phenomenological Studies\label{sec:pheno}}
We now turn to a detailed phenomenologically study of various heavy flavor observables
in longitudinally polarized DIS that are of potential interest for 
the physics program at a future EIC \cite{Boer:2011fh}. 
As in our previous study on the inclusive DIS structure function $g_1$ in Ref.~\cite{Hekhorn:2018ywm},
the main focus will be on the sensitivity of the corresponding
double-spin asymmetries to the
helicity gluon density $\Delta \Pg$ in the kinematic range $10^{-3}\lesssim x \lesssim 5\times 10^{-2}$ that
will be experimentally explored for the first time at the EIC.

\begin{figure}[ht!]
\begin{center}
\includegraphics[width=0.48\textwidth]{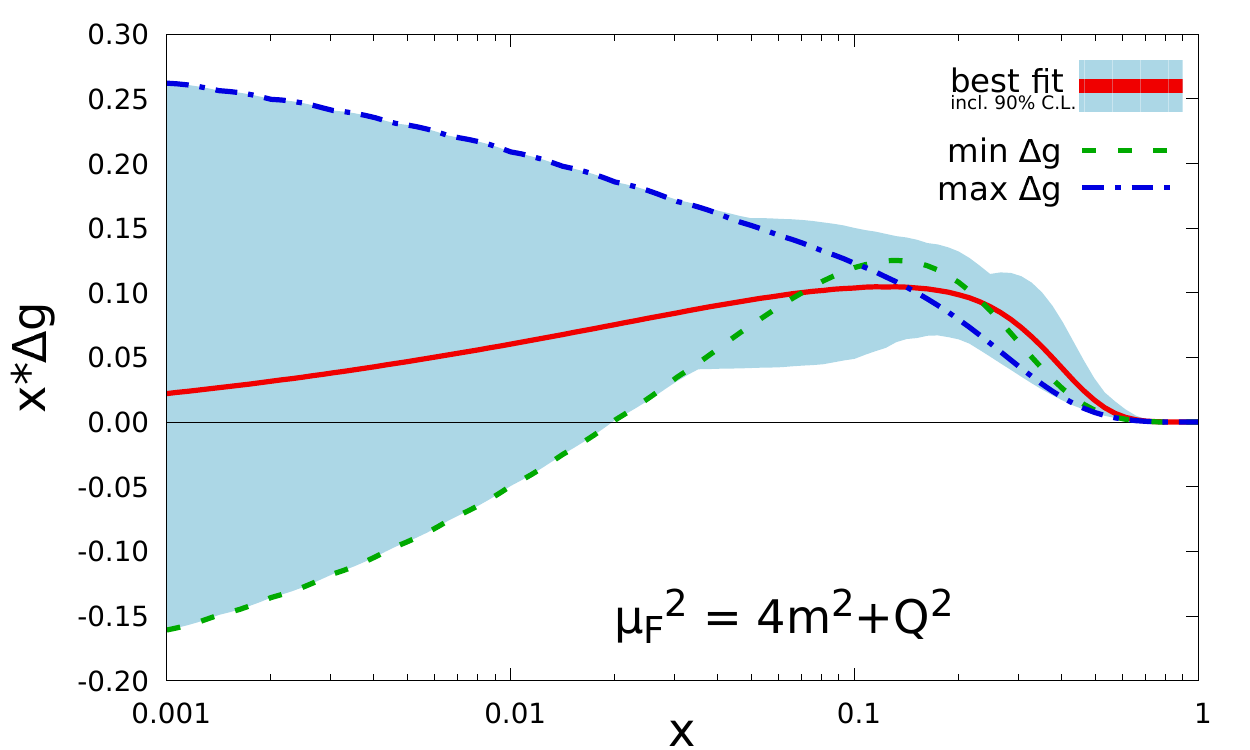}
\end{center}
\vspace*{-0.4cm}
\caption{The helicity gluon distribution $x\,\Delta g$ from the 
DSSV global analysis \cite{deFlorian:2008mr,deFlorian:2014yva}
as a function of $x$ for a typical scale
$\mu_F$ relevant for charm quark production in DIS at the EIC
($m=\SI{1.5}{\GeV}$, $Q^2=\SI{10}{\GeV^2}$).
The solid, dashed, and dot-dashed line represents the
best fit, the ``max $\Delta \Pg$'', and ``min $\Delta \Pg$'' set,
respectively. The shaded area is the DSSV uncertainty estimate for
$x\,\Delta g$ at $90\%$ C.L. \label{fig:gluon}}
\end{figure}
To this end, we adopt for all our numerical calculations 
the helicity PDFs of the DSSV group along with their uncertainty estimates 
\cite{deFlorian:2008mr,deFlorian:2014yva}. In particular, we will highlight the results
obtained by utilizing the two most extreme sets that provide the largest excursions 
from the DSSV best fit gluon PDF in the relevant small $x$ region. 
Throughout the remainder of the paper, these two sets are labeled 
as ``max $\Delta \Pg$'' and ``min $\Delta \Pg$'', respectively.
The DSSV best fit gluon PDF \cite{deFlorian:2008mr,deFlorian:2014yva} is shown in Fig.~\ref{fig:gluon} along
with the corresponding uncertainty estimate at $90\%$ confidence level (C.L.) and the two extreme sets.
As can be seen, in the small $x$-region of interest, the sets ``max $\Delta \Pg$'' and
``min $\Delta \Pg$'' form the envelope of the uncertainty band. This remains true
for all relevant scales $\mu_F$ used in our study.  It should be remarked that for those
members of the DSSV uncertainty band that exhibit a node at small $x$, the position of the node moves
to smaller values of the momentum fraction with increasing scale $\mu_F$.

Since the DSSV sets are only available at NLO accuracy, we have to use them also 
in all our calculations at LO accuracy and, hence, for the corresponding ``$K$-factors''.
To compute all the unpolarized DIS structure functions and HQ distributions
that appear in the denominator of the experimentally relevant double-spin asymmetries
to be defined below,
we adopt the NLO $90\%$ C.L.\ PDF set of the MSTW group \cite{Martin:2009iq}.
This set also serves as the reference set in the DSSV global analysis
in ensuring compliance with the positivity limit for helicity PDFs.
We also use the values of the strong coupling $\alpha_s(\mu_R)$ at NLO accuracy 
as determined in the MSTW fit \cite{Martin:2009iq}. The choice of the factorization
and renormalization scales, $\mu_F$ and $\mu_R$, respectively, 
depends on the HQ observable under consideration and,
hence, will be indicated in each case below. For all fully inclusive HQ distributions
we choose \cite{Hekhorn:2018ywm} $\mu_F^2=\mu_R^2=4m^2+Q^2$. We are only going to 
consider (anti-)charm quark electroproduction throughout this paper.

It should be also mentioned that all kinematic quantities such as transverse momenta,
rapidities, and angles of the observed heavy quark and/or antiquark
refer to the virtual photon-proton ($\Pggx\Pp$) c.m.s.\ frame, 
where the positive $z$-axis coincides with the momentum direction of the hadron.
The latter choice differs from the one taken in Ref.~\cite{Laenen:1992xs}, where the
$\Pggx$ travels in the $z$-direction. More details and explicit formulas related
to kinematical considerations comprising the choice of momenta, 
relevant Jacobians and expressions for the various integration limits can 
be found in Refs.~\cite{Laenen:1992xs,Harris:1995pr,ref:phd}.

\begin{figure}[th!]
\begin{center}
\includegraphics[width=0.49\textwidth]{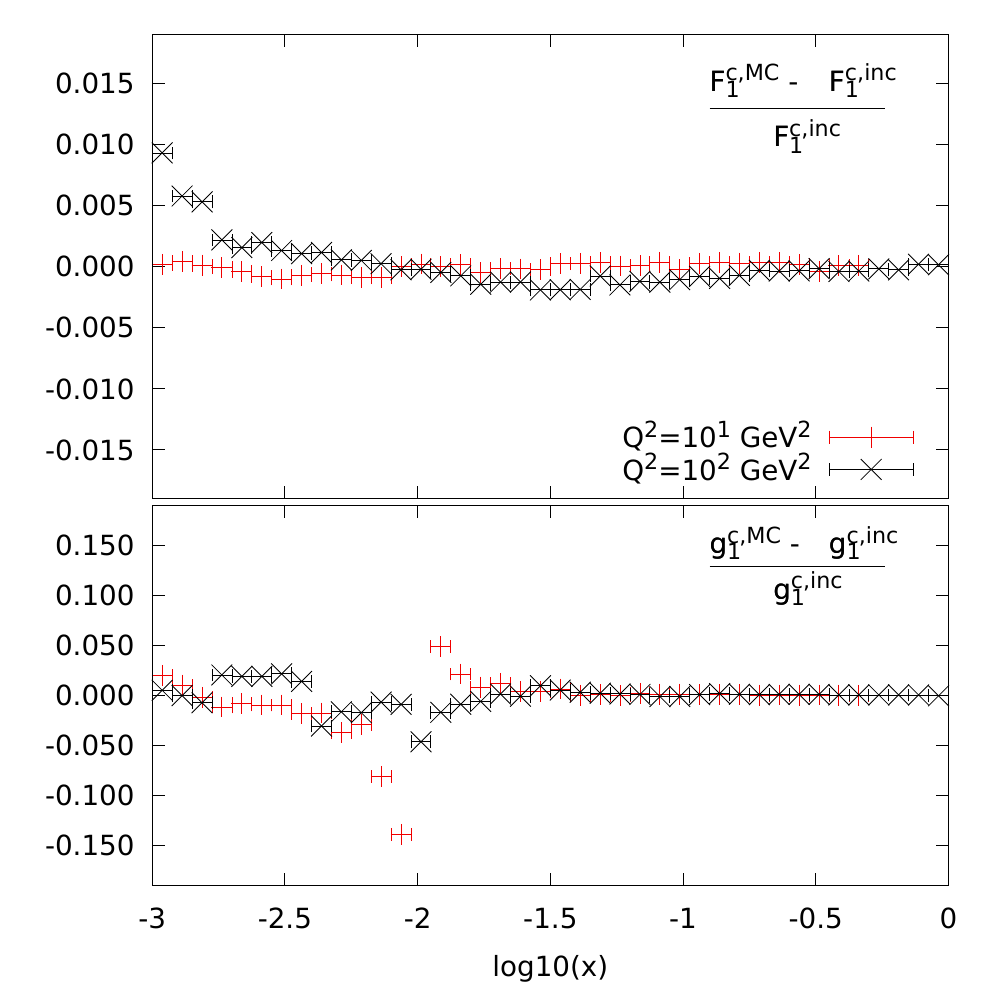}
\end{center}
\vspace*{-0.4cm}
\caption{Differences of the fully inclusive unpolarized (upper panel) and polarized (lower panel) 
charm structure functions obtained with the MC phase space integration, 
$F_1^{\Pqc,\mathrm{MC}}$ and $g_1^{\Pqc,\mathrm{MC}}$, and the largely analytical code of Ref.~\cite{Hekhorn:2018ywm},
$F_1^{\Pqc,\mathrm{inc}}$ and $g_1^{\Pqc,\mathrm{inc}}$. In both cases, the numerical difference 
is normalized to the respective analytical result. 
The comparison is presented as a function of $x$ and for two different values of $Q^2$.} 
\label{fig:relErrorF1g1}
\end{figure}
\begin{figure*}[th!]
\begin{center}
\includegraphics[width=0.48\textwidth]{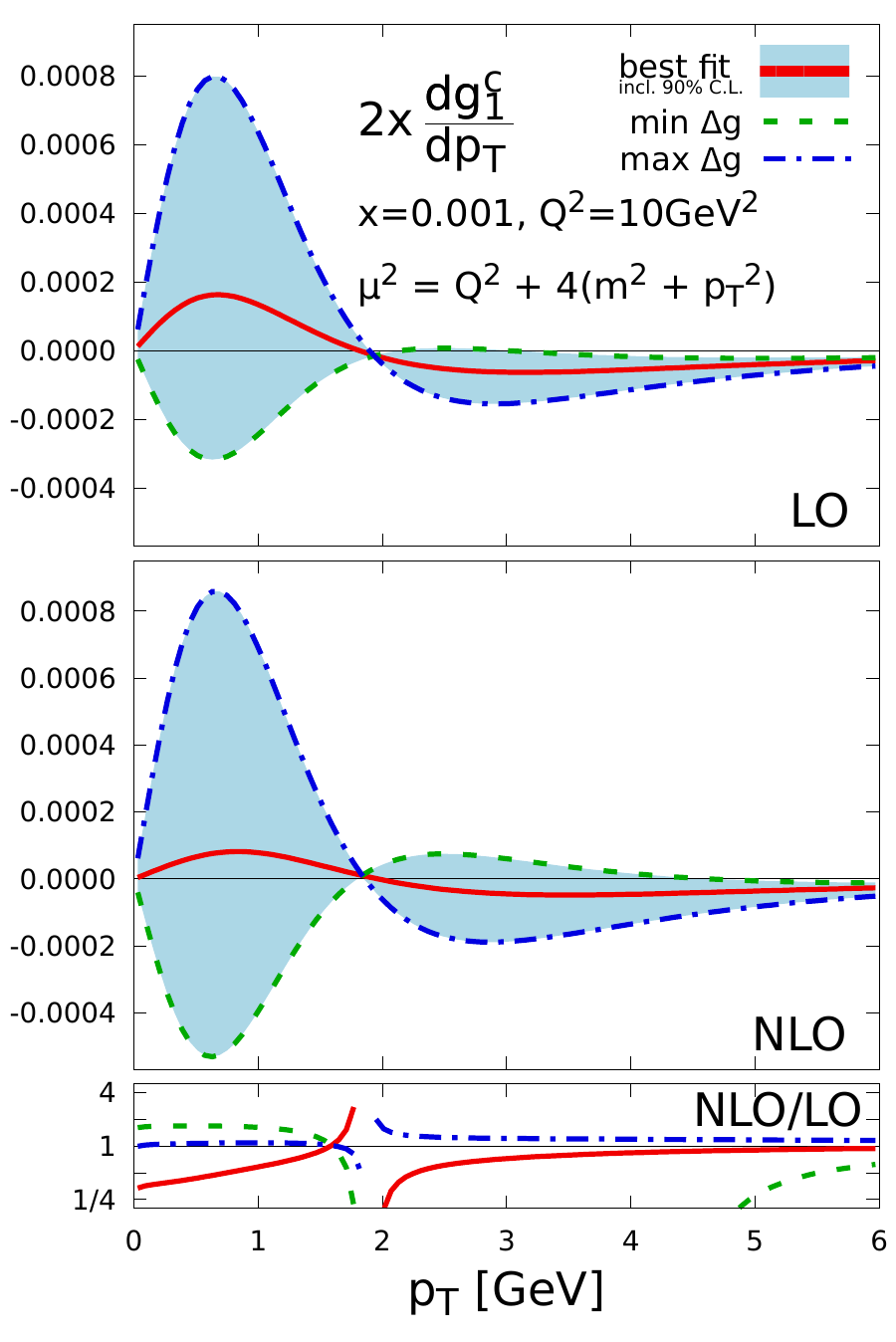}
\hfill
\includegraphics[width=0.48\textwidth]{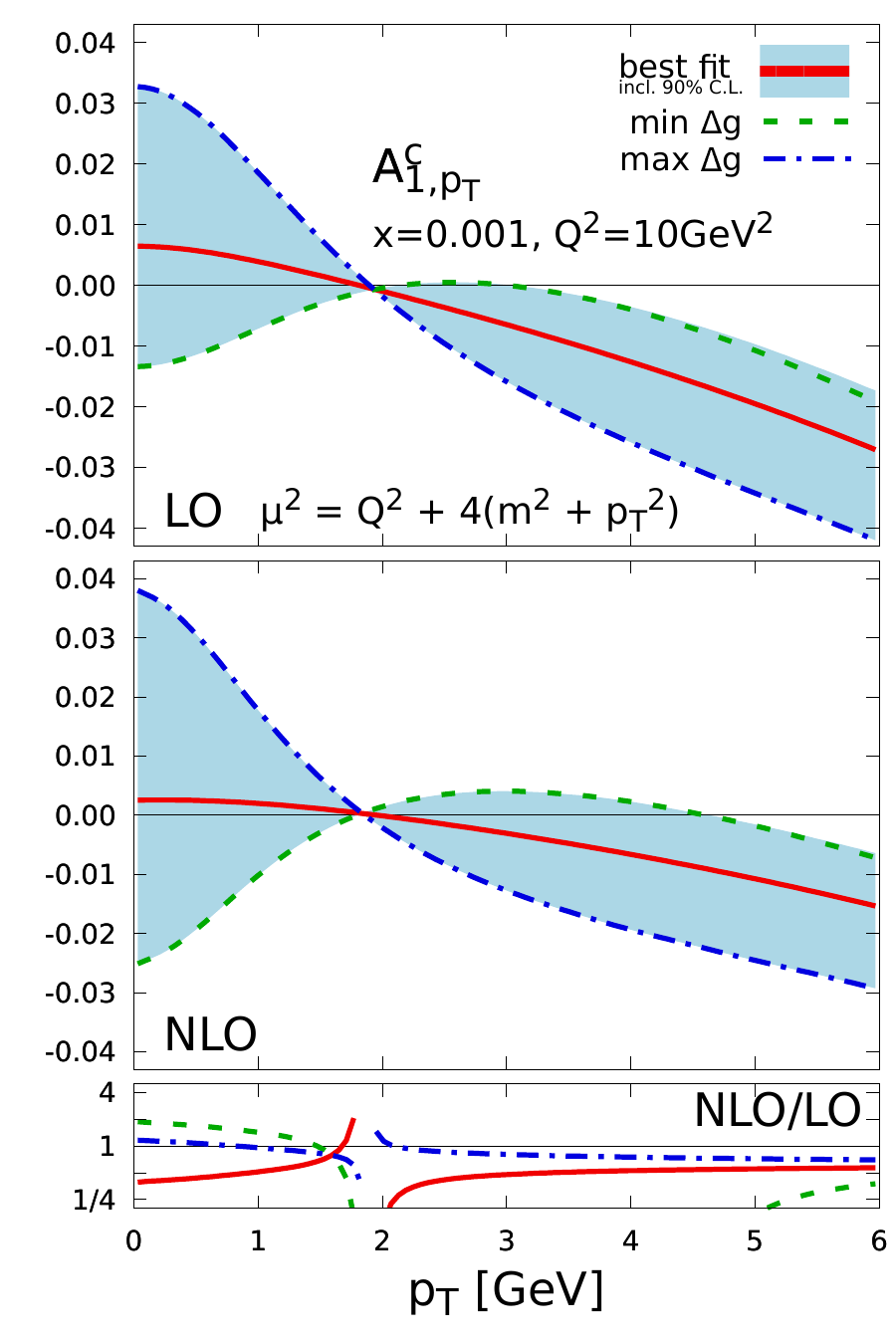}
\end{center}
\vspace*{-0.4cm}
\caption{The $p_T$-differential distribution $2x\,dg_1^{\Pqc}/dp_{T}$ (left-hand-side) and double-spin asymmetry
$A_{1,p_{T}}^{\Pqc}$ (right-hand-side) as a function of $p_T$ for fixed $x=0.001$ and $Q^2=\SI{10}{\GeV^2}$.
The LO and NLO results are given in the upper and middle panels, respectively, and the bottom panels show
the corresponding $K$-factors. The solid lines refer to the DSSV best fit $\Delta \Pg$ and the dashed and
dot-dashed lines are obtained with the ``min $\Delta \Pg$'' and ``max $\Delta \Pg$'' sets, respectively, 
spanning the uncertainty range (shaded bands). 
All results have been obtained with the choice $\mu^2=\mu_F^2 = \mu_R^2 = Q^2 + 4(m^2+p_T^2)$.}
\label{fig:dg1dpt}
\end{figure*}
As a first important step, we validate our newly developed parton-level MC code 
against the results already known from our previous calculations based on
largely analytical methods \cite{Hekhorn:2018ywm}. 
Such a comparison is possible for any single-inclusive HQ observable in DIS
that can be expressed in terms of the transverse momentum $p_T$ or 
transverse mass $m_T=\sqrt{p_T^2+m^2}$ and the
rapidity $y$ of the observed heavy (anti)quark; 
see, e.g., the appendices in Refs.~\cite{Laenen:1992xs,ref:phd} for 
details concerning the appropriate Jacobians for the required changes of integration variables.
As an example, we consider in Fig.~\ref{fig:relErrorF1g1} the fully inclusive 
charm contributions $F_1^{\Pqc}$ and $g_1^{\Pqc}$ 
to the corresponding unpolarized and polarized DIS structure functions $F_1$ and $g_1$, respectively,
which have been studied extensively in \cite{Hekhorn:2018ywm}.
The upper and lower panels of Fig.~\ref{fig:relErrorF1g1} show the relative differences 
of the numerical results for $F_1^{\Pqc}$ and $g_1^{\Pqc}$ 
obtained with the largely analytical code of Ref.~\cite{Hekhorn:2018ywm}, labeled as ``inc'', 
and the MC method described in this paper. Results are presented as a function of $x$ for
two different values of $Q^2$.
We shall note that the necessary multi-dimensional MC integrations 
are performed with an enhanced adaptive {\sc Vegas} routine, see Ref.~\cite{Kauer:2001sp}.
To obtain numerically stable results, it suffices to use 500k points and 
5 iterations at LO accuracy and 4M points and 20 iterations 
for all computations involving NLO corrections. 

As can be seen, the numerical differences for $F_1^{\Pqc}$ are 
at the sub-percent level throughout and for most of the kinematic range shown
in Fig.~\ref{fig:relErrorF1g1} even close to one per-mille. 
Differences for $g_1^{\Pqc}$ are somewhat larger, but usually better than $1-2\%$, except in
the vicinity of nodes where $g_1^{\Pqc}$ is numerically very small and changes sign.
The comparisons in Fig.~\ref{fig:relErrorF1g1}
reflect the typical numerical accuracy that can be achieved with the chosen number
of points and iterations in the {\sc Vegas} integration. 
In general, for any single-inclusive HQ observable, see Ref.~\cite{ref:phd} for more examples,
the integrations leading to the unpolarized result 
are, as expected, noticeably more stable than the corresponding ones for the polarized distributions.
The latter may exhibit nodes and, in addition, 
suffers from significantly larger numerical cancellations 
between the different contributions to the full NLO result. We wish to stress here, that we have
validated our MC code also against all the single-inclusive HQ distributions shown below,  
with similar numerical differences to the largely analytical code 
as documented in Fig.~\ref{fig:relErrorF1g1}.

\begin{figure*}[th!]
\begin{center}
\includegraphics[width=0.48\textwidth]{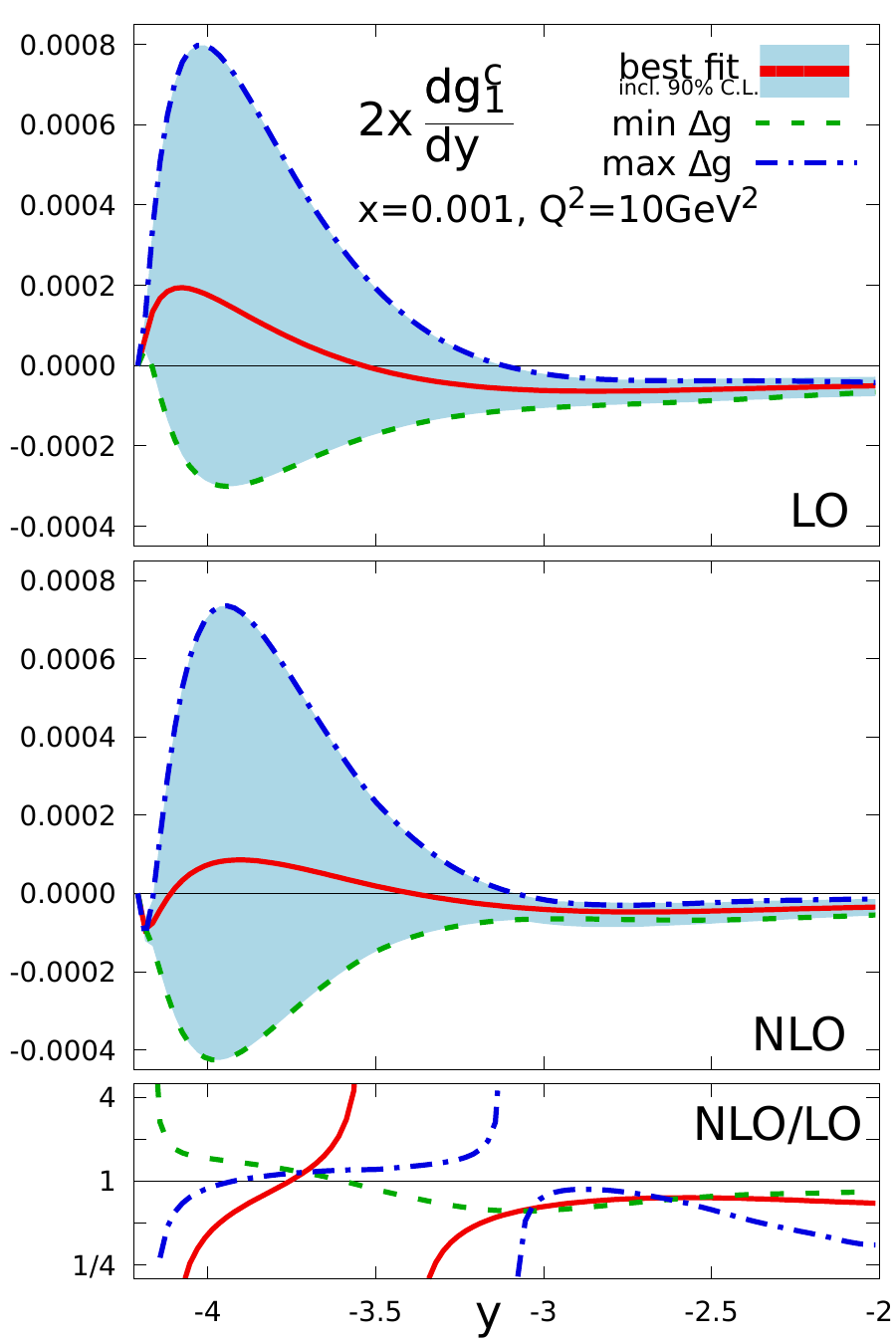}
\hfill
\includegraphics[width=0.48\textwidth]{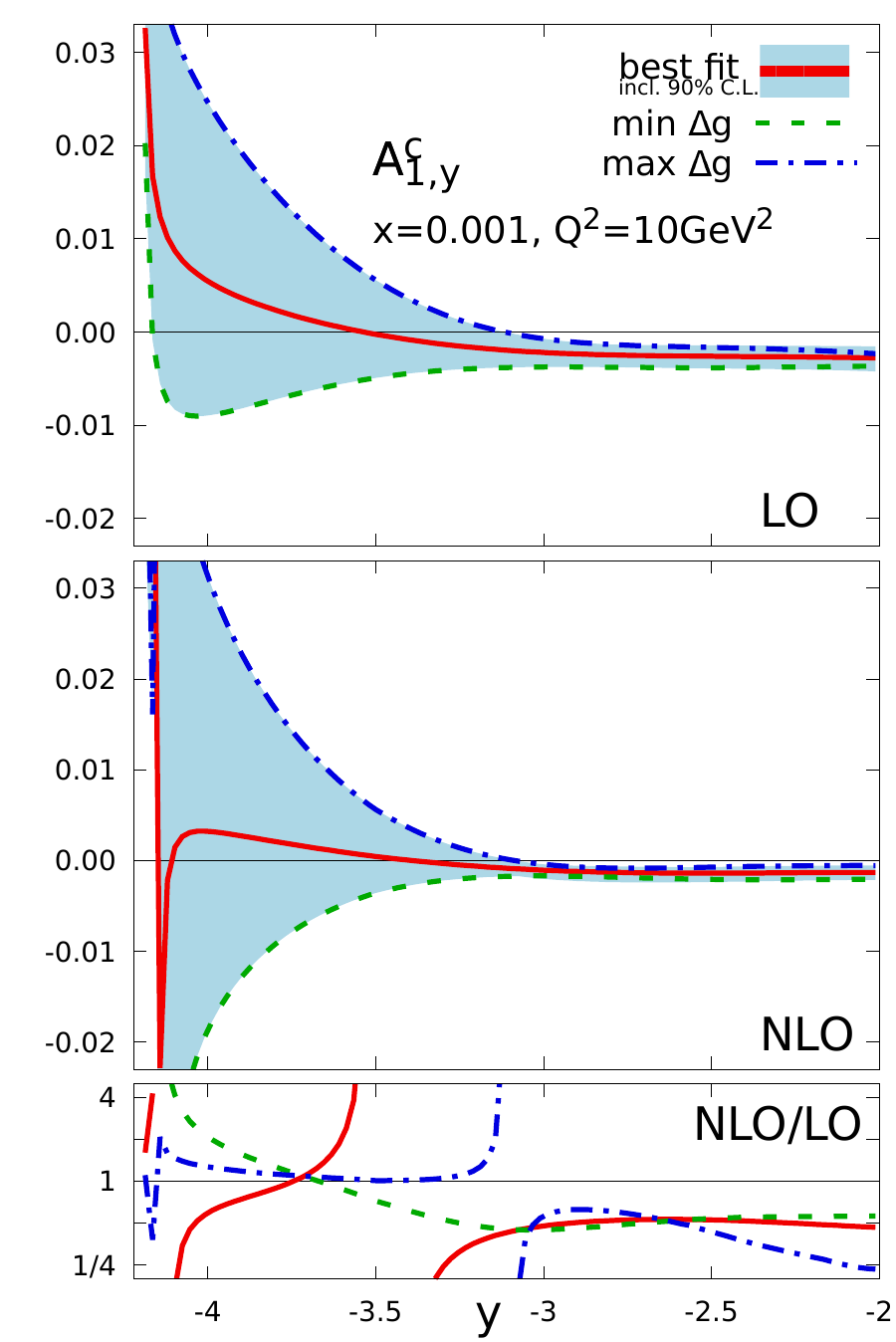}
\end{center}
\vspace*{-0.4cm}
\caption{As in Fig.~\ref{fig:dg1dpt} but now for the rapidity distribution $2x\,dg^{\Pqc}_1/dy$ (left-hand-side)
and double-spin asymmetry $A^{\Pqc}_{1,y}$ (right-hand-side).} 
\label{fig:dg1dy}
\end{figure*}
After this prelude, we shall investigate the $p_T$- and $y$-differential single-inclusive distributions
of the longitudinally polarized DIS charm structure function $g_1^{\Pqc}$ in Figs.~\ref{fig:dg1dpt} and
\ref{fig:dg1dy}, respectively. In both cases we choose a fixed momentum fraction 
$x=0.001$ and set $Q^2=\SI{10}{\GeV^2}$,
which are both well within the interesting kinematic range to be explored for the first time at a future EIC.
As was mentioned already, the results in this paper have been obtained for a detected heavy anticharm quark.
We note that there is in fact a very small asymmetry between observing the heavy quark or the heavy antiquark
distributions at NLO accuracy \cite{Laenen:1992zk,Hekhorn:2018ywm}, 
which is, however, unlikely to be of any phenomenological relevance at the EIC 
in case of polarized DIS.
For all unpolarized $p_T$- and $y$-differential distributions that appear in the
denominator of the relevant double-spin asymmetries, see Eq.~(\ref{eq:a1def}) below, we have obtained excellent numerically agreement
with the results given in \cite{Laenen:1992xs} to which we refer the interested reader for 
a discussion of the most important findings in the unpolarized case. Of course, for all our purposes we have 
updated the relevant results in Ref.~\cite{Laenen:1992xs} using the MSTW set of PDFs and conventions for 
the strong coupling, but we refrain from reproducing them here.

Figure~\ref{fig:dg1dpt} shows $2x\,dg_1^{\Pqc}/dp_T$ (left-hand-side) and the corresponding double-spin
asymmetry 
\begin{equation}
A_{1,p_{T}}^{\Pqc} \equiv \frac{dg_1^{\Pqc}/dp_T}{dF_1^{\Pqc}/dp_T}
\label{eq:a1def}
\end{equation}
(right-hand-side). The upper and middle row refer to the calculations
at LO and NLO accuracy, respectively, and the bottom panels illustrate the $K$-factor,
i.e., the ratio of the NLO and LO results, as a function of $p_T$. 
We have restricted $p_T$ to range where measurements appear to be conceivable at the EIC 
for the given values of $x$ and $Q^2$, and rapidity is integrated 
over the entire phase space available. In any case,
small values of $p_T$ correspond to probing $\Delta \Pg$ at the smallest momentum fractions kinematically allowed,
which is one of the main goals of the EIC spin program. More precisely, for the typical kinematics, $x$ and $Q^2$,
selected in Fig.~\ref{fig:dg1dpt}, the values of momentum fraction in $\Delta \Pg$ 
that are predominantly probed in the convolutional integral with the hard coefficient functions is
approximately in the range between $0.015$ and $0.05$, i.e., about a decade larger than the
chosen value of $x$;
see the appendix of Ref.~\cite{ref:phd} for a collection of all relevant equations entering
in this estimate.
In addition, the limitation to moderate values of $p_T$ has the extra benefit 
that one does not have to worry about any potentially large logarithms of the type $\ln (p_T/m)$ 
and their eventual resummation to all orders in perturbation theory at this stage. This might
be an interesting subject to pursue in the future, 
following the work already available in the unpolarized case \cite{Cacciari:1993mq}. 

In contrast to the fully inclusive calculation shown in Fig.~\ref{fig:relErrorF1g1}, 
and following Ref.~\cite{Laenen:1992xs}, we now choose for all single-inclusive calculations
the scale $\mu^2=\mu_F^2=\mu_R^2=Q^2 + 4(m^2 + p_T^2)$ to account for the presence of the additional
hard momentum $p_T$. If one were to choose a $p_T$-independent scale, such as $\mu^2= Q^2 + 4m^2$, 
in the calculation, one would observe even larger QCD corrections, in particular, 
towards higher values of $p_T$, destabilizing the
perturbative expansion; a similar observation was made in the unpolarized case, see Ref.~\cite{Laenen:1992zk}.
Below, we shall discuss in more detail the impact and relevance of scale variations for $dg_1^{\Pqc}/dp_T$.

From the panels in Fig.~\ref{fig:dg1dpt} one can infer that all gluon distributions $\Delta \Pg$ that lie
within the uncertainty estimate provided by DSSV (shaded bands) produce a node in the $p_T$-distribution both
at LO and NLO accuracy somewhere in the range 
$p_T\in \SIlist[list-units=brackets,list-pair-separator={,\,}]{1.5;2.0}{\GeV}$, naturally accompanied by
large QCD corrections in their vicinity. The feature of the common node 
has no simple kinematic explanation but is deeply rooted in the convolutional
integral of the PDFs with the oscillating coefficient functions and their functional properties; 
see Ref.~\cite{Hekhorn:2018ywm}.
In the given range of $p_T$, where the $\Pggx\Pg$-subprocess
dominates, the two extreme sets min and max $\Delta \Pg$ highlighted in Fig.~\ref{fig:gluon}
also provide the envelope of the $p_T$--differential structure function. 
In general, a large positive $\Delta \Pg$ leads to the smallest QCD corrections 
as can be seen best from the bottom panels. However,
as was already noted in case of the fully inclusive charm
structure function $g_1^{\Pqc}$ in Ref.~\cite{Hekhorn:2018ywm}, the NLO corrections are substantial throughout
and should be included in any future global QCD analysis including HQ electroproduction data.
Since the NLO corrections strongly depend on $\Delta \Pg$ and $p_T$ one must not resort to any approximations,
such as a constant $K$-factor in future fits.

The experimentally relevant $A_{1,p_{T}}^{\Pqc}$ is larger 
than in the fully inclusive case, see Ref.~\cite{Hekhorn:2018ywm},
and can amount to a few percent. Even an initial measurement 
with a absolute accuracy of about $0.5\%$ at the EIC would prove to be very
valuable in determining the $x$-shape of $\Delta \Pg$ more precisely. 
This would be true, in particular, if such a measurement
could be carried out for at least two different bins of $p_T$, ideally to the left and to the right of the 
node in $A_{1,p_{T}}^{\Pqc}$ to verify if the sign change predicted by all members of the DSSV uncertainty
band for $\Delta \Pg$ is indeed realized or not.

\begin{figure}[ht!]
\begin{center}
\includegraphics[width=0.48\textwidth]{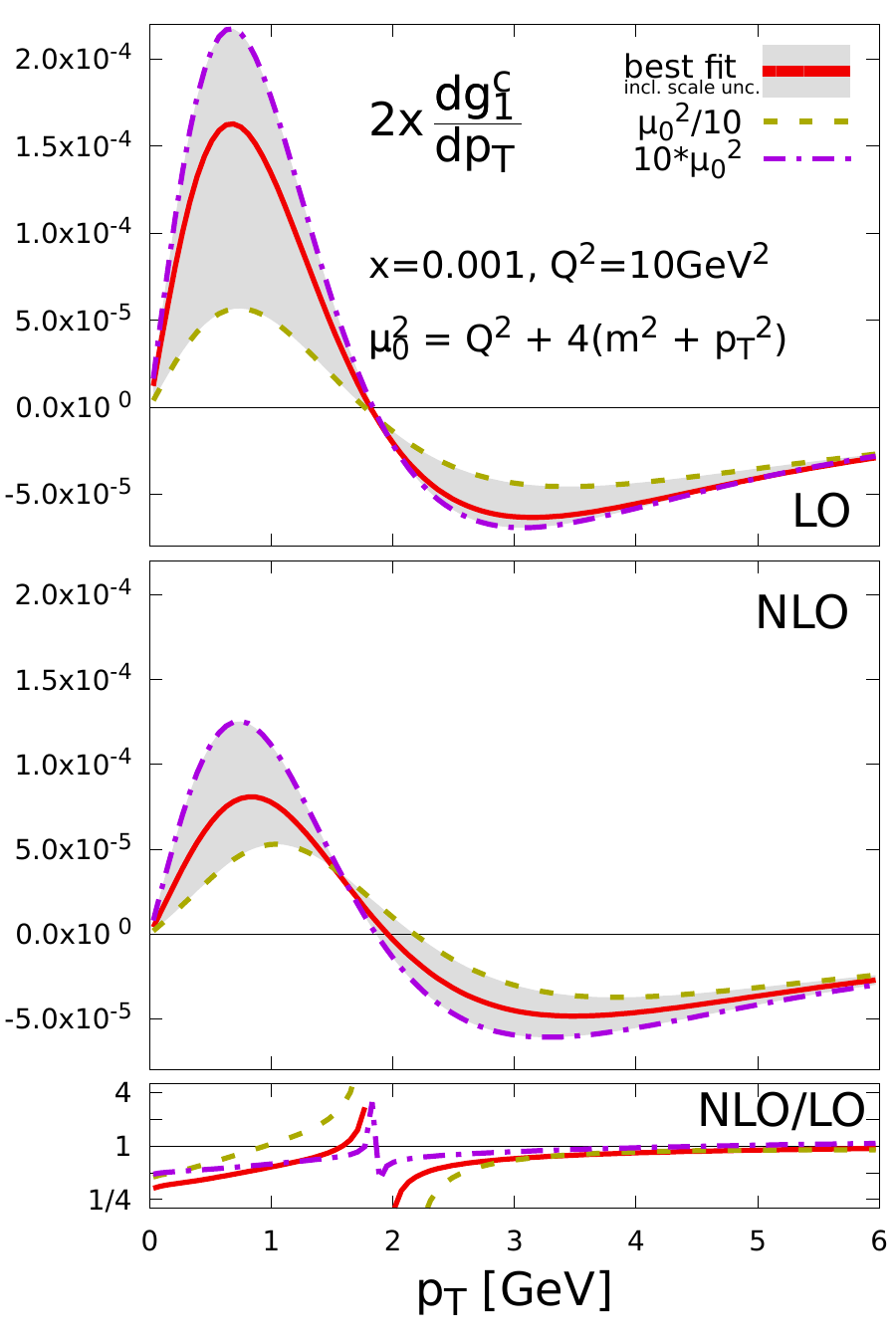}
\end{center}
\vspace*{-0.4cm}
\caption{Dependence of $2x\,g_1^{\Pqc}/dp_T$ at $x=0.001$ 
calculated with the DSSV best fit (solid lines)
on simultaneous variations (shaded bands) of $\mu^2=\mu_F^2=\mu_R^2$ in the range
$\mu^2\in \{10\mu_0^2,\mu_0^2/10\}$ where $\mu_0^2 = Q^2 + 4(m^2+p_T^2)$
and $Q^2=\SI{10}{\GeV^2}$. The top and middle panel refers to
results at LO and NLO accuracy, respectively.
The bottom panel shows the scale dependence of the $K$-factor.}
\label{fig:dg1dpt-scale1}
\end{figure}
\begin{figure*}[th!]
\begin{center}
\includegraphics[width=0.48\textwidth]{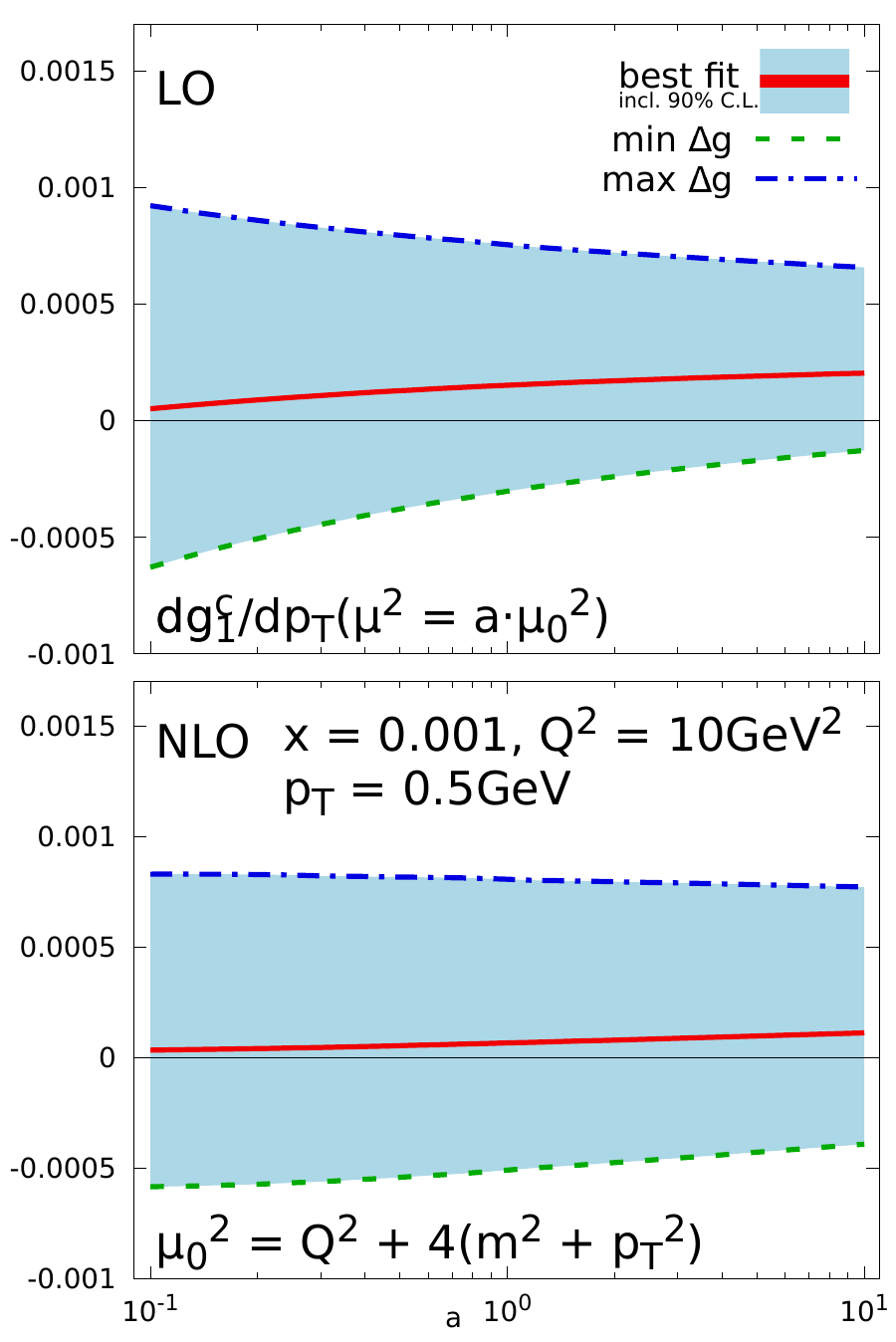}
\hfill
\includegraphics[width=0.48\textwidth]{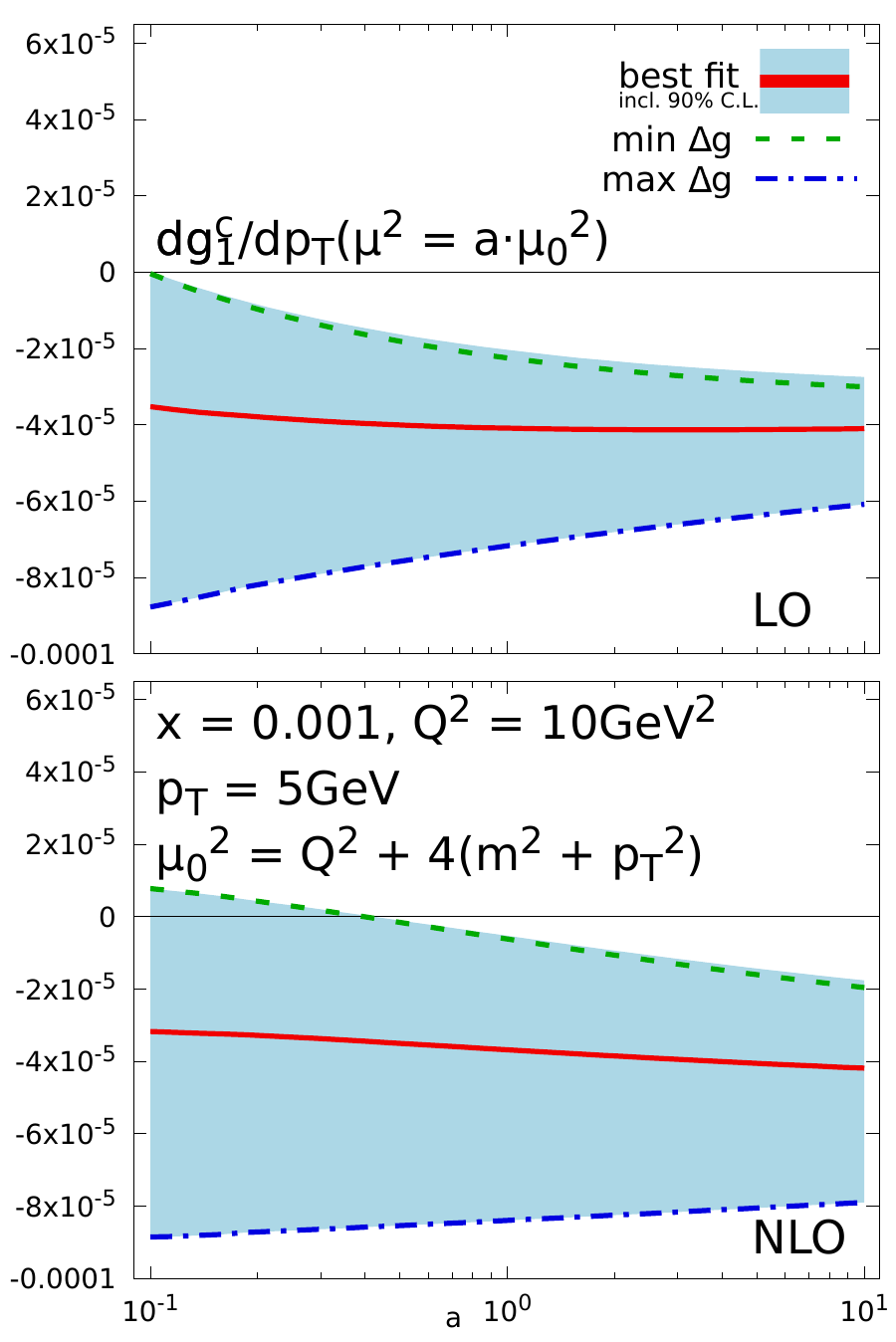}
\end{center}
\vspace*{-0.4cm}
\caption{Scale variation of $dg^{\Pqc}_1/dp_T(\mu^2)$ in the range $\mu^2=a\cdot\mu_0^2$  
where $a\in[0.1,10]$ and $\mu_0^2=Q^2 + 4(m^2+p_T)^2$. The results are obtained at LO and NLO accuracy,
upper and lower row, respectively, for $x=0.001$, $Q^2=\SI{10}{\GeV^2}$
and two fixed values of $p_T$: $\SI{0.5}{\GeV}$ (left-hand-side) and $\SI{5}{\GeV}$ (right-hand-side).
The shaded bands represent the entire suite of DSSV helicity PDF uncertainty sets; highlighted are the 
best fit (solid line), min $\Delta \Pg$ (dashed line), and max $\Delta \Pg$ (dot-dashed line) sets.}
\label{fig:dg1dpt-scale2}
\end{figure*}
Analogously, Fig.~\ref{fig:dg1dy} presents another single-inclusive distribution of potential
interest for the physics program at the EIC: $2x\,dg_1^{\Pqc}/dy$ (left-hand-side). 
As before, the right-hand-side of the plot 
gives the corresponding double-spin asymmetry $A^{\Pqc}_{1,y}$, defined analogously to Eq.~(\ref{eq:a1def}), 
and the DIS kinematics is again fixed to $x=0.001$ and $Q^2=\SI{10}{\GeV^2}$. The transverse momentum of
the observed charm antiquark is integrated over the entire phase phase, and the rapidity $y$ is
defined in the $\Pggx\Pp$ c.m.s.\ frame with negative $y$ pointing in the direction in which the
$\Pggx$ travels. We restrict ourselves to the region of $-4\lesssim y \lesssim -2.5$ 
where $\Delta \Pg$ is probed at the smallest momentum fractions for any given $p_T$.
As for the $p_T$-distribution, different $\Delta \Pg$ from the DSSV uncertainty estimate lead
to different predictions for $2x\,dg_1^{\Pqc}/dy$ which all peak at around $y\simeq -4$ though. Here, the
double-spin asymmetry $A^{\Pqc}_{1,y}$ can again reach values at the percent-level and should be measurable at the EIC.
At higher values of $y$, the differences in $A^{\Pqc}_{1,y}$ quickly diminish and 
within the DSSV uncertainty estimate a small negative $A^{\Pqc}_{1,y}$ is expected.
Given the anticipated unprecedented, high luminosity of the EIC, it might be even worth studying 
double-differential distributions in both $p_T$ and $y$ for some suitable binning in the future to
further optimize the sensitivity of HQ DIS observables to $\Delta \Pg$. 
We note that the somewhat erratic behavior at NLO observed for $y<-4$ is due to 
end-of-phase space effects, i.e., in this region only the emission of soft gluons 
is kinematically possible.
Corresponding large logarithmic terms would need to be resummed to all orders in
perturbation theory to stabilize the result. A similar observation was made in
the unpolarized case, see Ref.~\cite{Laenen:1992zk}.

The combined set of the next two plots, Figs.~\ref{fig:dg1dpt-scale1} and \ref{fig:dg1dpt-scale2}, 
gives a flavor of the remaining scale dependence of the 
single-inclusive distribution $dg_1^{\Pqc}/dp_T$ discussed in Fig.~\ref{fig:dg1dpt}, which we deem to be
the phenomenologically most important one. More specifically, 
Fig.~\ref{fig:dg1dpt-scale1} illustrates the dependence of $2x\,g_1^{\Pqc}/dp_T$ 
obtained with the DSSV best fit (solid lines) at LO (top panel) and NLO (middle panel) accuracy under
simultaneous variations (shaded bands) of $\mu^2=\mu_F^2=\mu_R^2$ in the range
$\mu^2\in \{10\mu_0^2,\mu_0^2/10\}$ where $\mu_0^2 = Q^2 + 4(m^2+p_T^2)$.
As can be seen, the variations for the best fit $\Delta \Pg$ are sizable but still significantly 
smaller than the spread in $dg_1^{\Pqc}/dp_T$ introduced by choosing the min or max $\Delta \Pg$ uncertainty sets,
c.f.\ Fig.~\ref{fig:dg1dpt}. Also the variations in the $K$-factor, given in the
bottom panel of Fig.~\ref{fig:dg1dpt-scale1}, are rather moderate, perhaps except for the
vicinity of the node in the $dg_1^{\Pqc}/dp_T$ distribution. At small $p_T$, near the peak of the
distribution, the scale variations are significantly smaller then the NLO corrections, as one
would hope for. At higher $p_T$ the width of the band is, however, rather similar in LO and
NLO accuracy.

\begin{figure*}[th!]
\begin{center}
\includegraphics[width=0.48\textwidth]{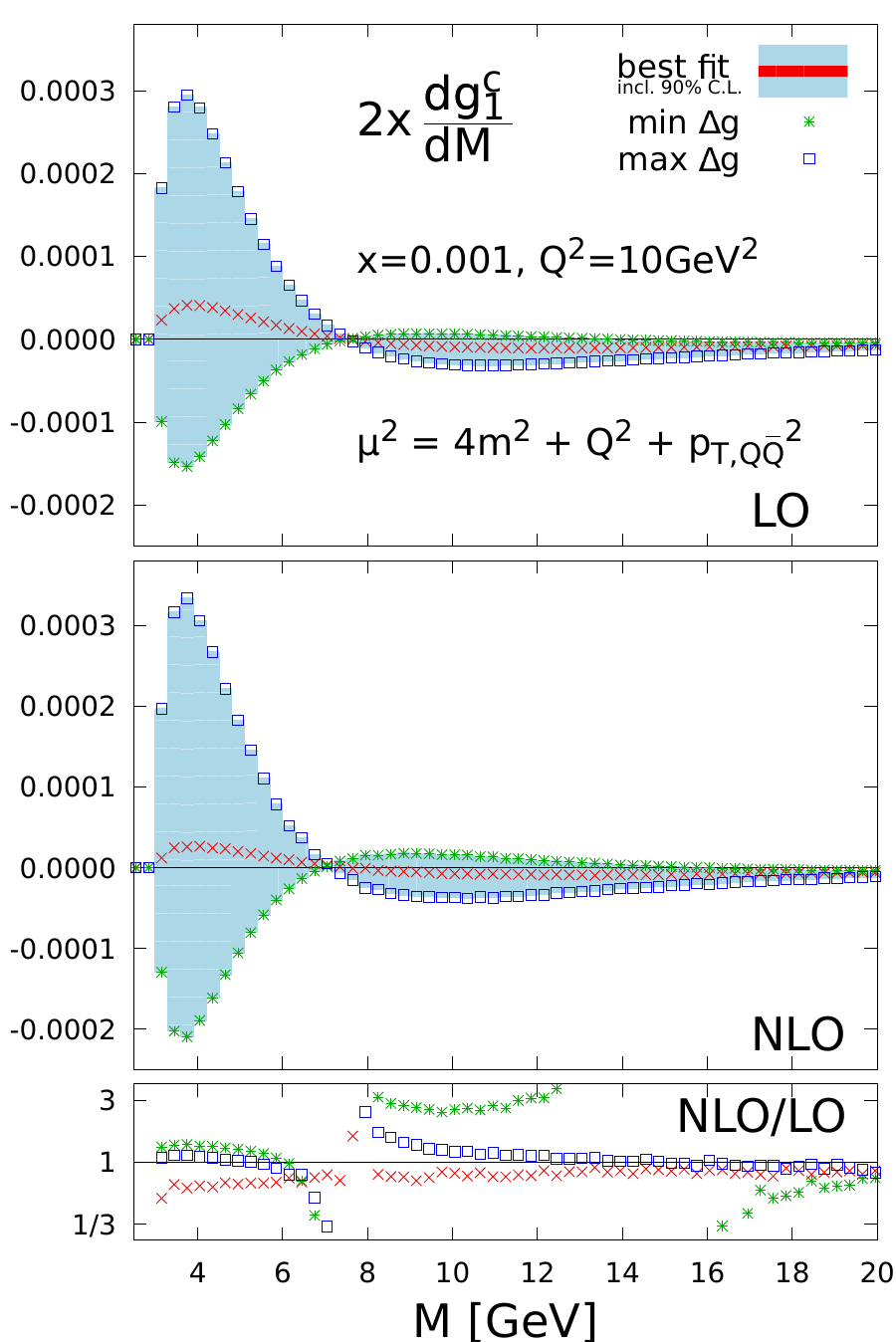}
\hfill
\includegraphics[width=0.48\textwidth]{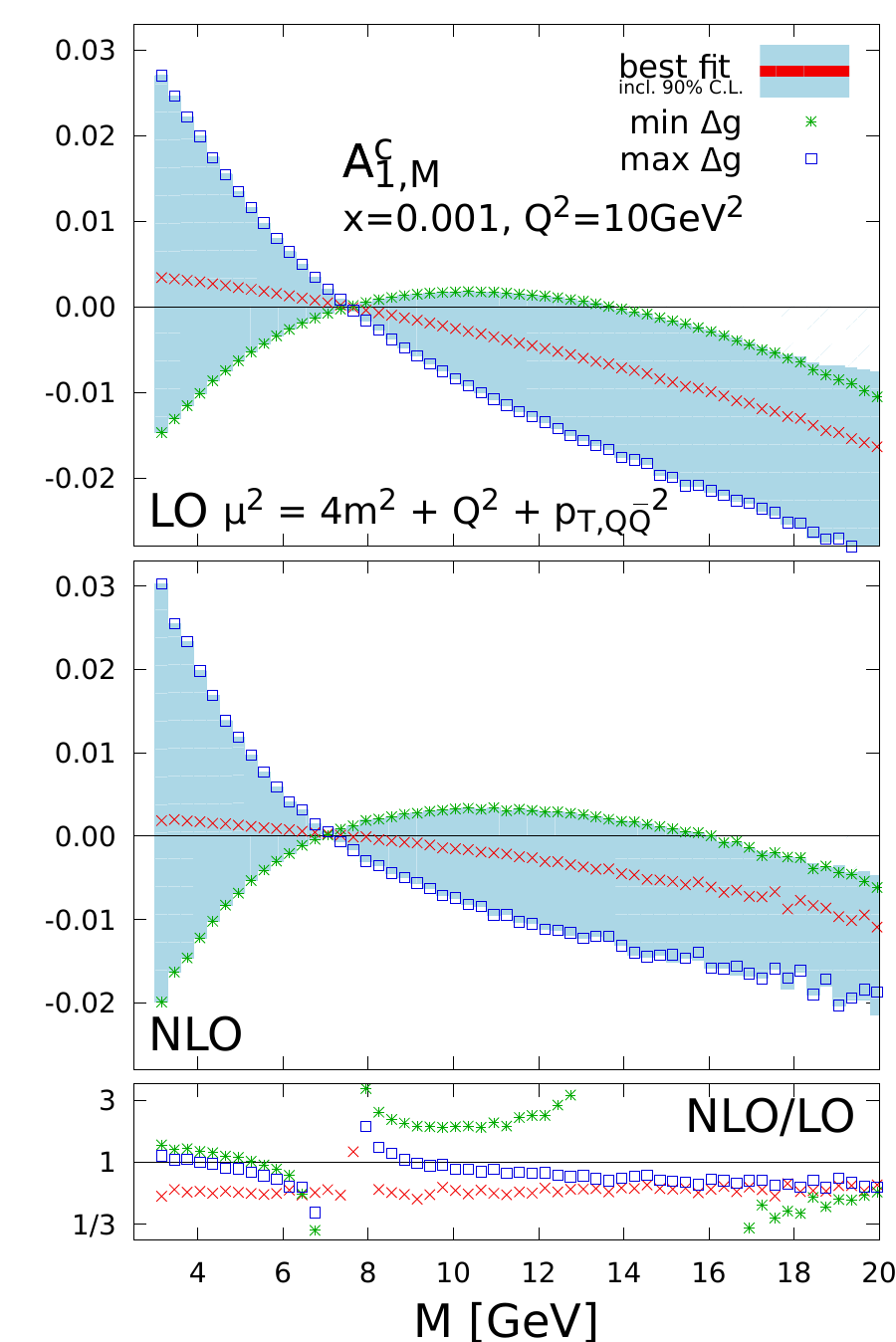}
\end{center}
\vspace*{-0.4cm}
\caption{As in Fig.~\ref{fig:dg1dpt} but now for the invariant mass distribution $2x\,dg_1^{\Pqc}/dM$ 
of the produced charm-anticharm pair (left-hand-side) and the corresponding
double-spin asymmetry $A^{\Pqc}_{1,M}$.} 
\label{fig:dg1dm}
\end{figure*}
Figure~\ref{fig:dg1dpt-scale2} compares the scale variations of $dg_1^{\Pqc}/dp_T$ for
the entire suite of DSSV helicity PDF uncertainty sets for two fixed values of $p_T$.
Again, the scale dependence is explored in the range $\mu^2=a\cdot\mu_0^2$ where $a\in[0.1,10]$
at LO (top) and NLO (bottom) accuracy. The shaded bands represent the range of the DSSV uncertainty estimate
with the best fit (solid line), min $\Delta \Pg$ (dashed line), and max $\Delta \Pg$ (dot-dashed line) sets
particularly highlighted. The panels on the left-hand-side are for $p_T=0.5\,\mathrm{GeV}$ 
and those on the right-hand-side correspond to setting $p_T=5\,\mathrm{GeV}$, 
i.e., we have picked two representative $p_T$-values to
the left and to the right of the node in $dg^{\Pqc}_1/dp_T$, c.f.\ Fig.~\ref{fig:dg1dpt}.
Most importantly, the extreme min and max $\Delta \Pg$ sets still provide the envelope of the 
DSSV predictions for $dg_1^{\Pqc}/dp_T$ in the entire range of $\mu$.
For any choice of scale, there is a clear, roughly constant sensitivity 
to $\Delta \Pg$ as discussed above.

In the remainder of this section 
we turn to observables which involve the detection of both the heavy quark {\em and} the heavy antiquark
and, hence, require the use of our MC code in their numerical evaluation. 
Again, we restrict ourselves to charm flavored (anti)quarks.
Presumably the most promising quantity is
the invariant mass distribution $2x\,dg_1^{\Pqc}/dM$ of the HQ pair 
which is shown in Fig.~\ref{fig:dg1dm} (left-hand-side) along with the corresponding
double-spin asymmetry $A^{\Pqc}_{1,M}$ defined in analogy to Eq.~(\ref{eq:a1def}). 
As our default choice of scale we now choose \cite{Harris:1995pr}
$\mu^2= 4m^2+Q^2 + p_{T,\PQ\PaQ}^2$, where $p_{T,\PQ\PaQ}$ denotes the combined transverse momentum of the 
produced heavy quark and antiquark pair. In addition to the results for the longitudinally polarized structure function
shown in Fig.~\ref{fig:dg1dm}, we have again successfully reproduced the results for the corresponding unpolarized
quantities given in Ref.~\cite{Harris:1995pr} to which we also refer the reader for a discussion of their relevant properties.

As before, we select $x=0.001$ and $Q^2=\SI{10}{\GeV^2}$ to define the DIS kinematics accessible at the EIC. 
The top and middle panels
of Fig.~\ref{fig:dg1dm} show the numerical results obtained at LO and NLO accuracy, respectively. The histograms
(denoted by different symbols) refer to our usual selection of DSSV sets (best fit, min $\Delta \Pg$, and max $\Delta \Pg$), and
the shaded bands correspond to the entire suite of DSSV uncertainty estimates. 
The lower panels give the respective $K$-factors
for $2x\,dg_1^{\Pqc}/dM$ and $A^{\Pqc}_{1,M}$. In general, we observe a rather similar behavior as for the single-inclusive
$p_T$-distributions shown in Fig.~\ref{fig:dg1dpt}. All results exhibit 
a node at around $M\simeq \SI{7}{\GeV}$, which is accompanied by large QCD corrections in its vicinity.
The asymmetries peak at around $M\simeq \SI{4}{\GeV}$. Again, initially a set of two measurements at the EIC,
one to the left and one to the right of the predicted node, would reveal important new insights into
the $x$-shape of $\Delta \Pg$. As before, all future measurements 
are required to resolve at least percent-level double-spin asymmetries to be of phenomenological relevance.

\begin{figure}[!]
\begin{center}
\includegraphics[width=0.48\textwidth]{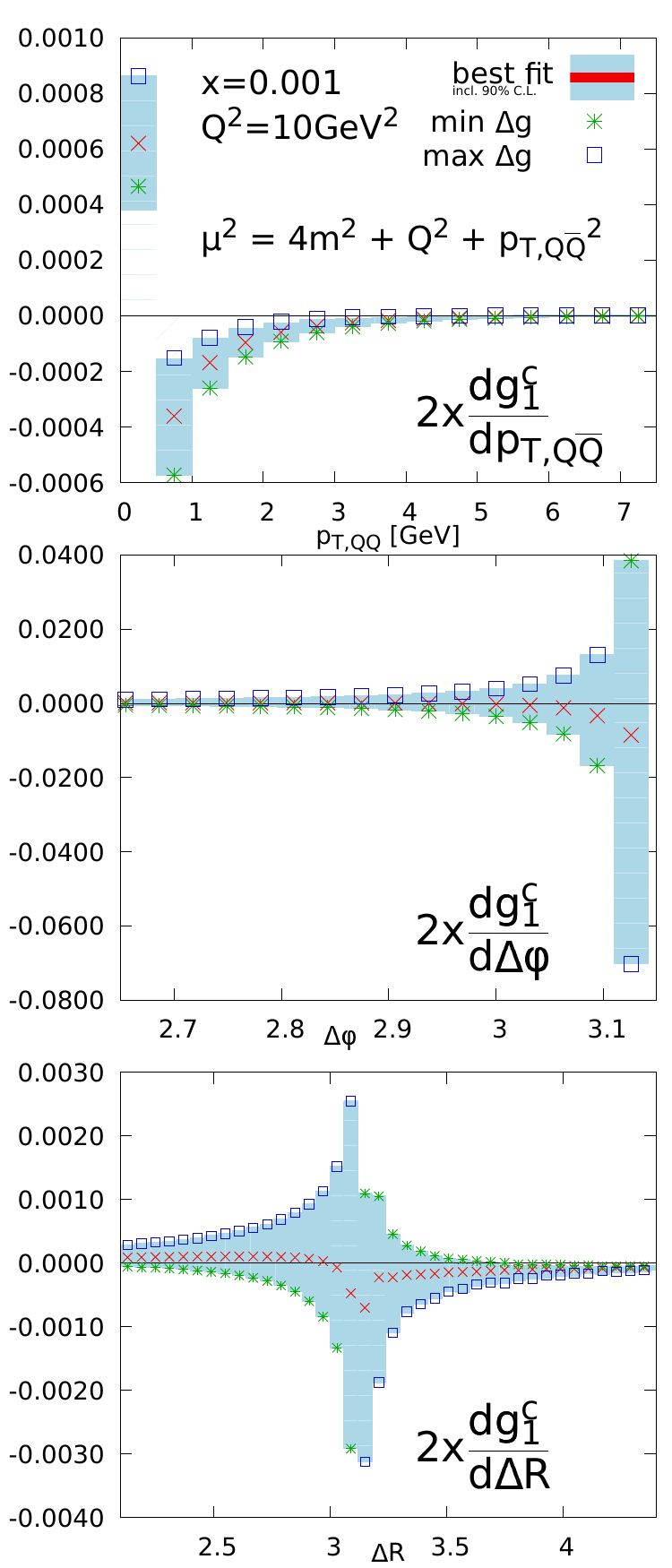}
\end{center}
\vspace*{-0.4cm}
\caption{The helicity DIS charm structure function $g_1^{\Pqc}$ differential in the transverse momentum
$p_{T,\PQ\PaQ}$ (top panel), the difference in azimuthal angle $\Delta \varphi$ (middle panel),
and HQ cone size variable $\Delta R$ (bottom panel), see text, of the produced charm-anticharm pair.
The histograms are obtained at NLO accuracy 
for the DSSV best fit, min $\Delta \Pg$, and max $\Delta \Pg$ sets of PDFs. The shaded bands
indicate the DSSV uncertainty estimates. All calculations are performed for $x=0.001$,
$Q^2=\SI{10}{\GeV^2}$, and $\mu^2= 4m^2+Q^2 + p_{T,\PQ\PaQ}^2$.} \label{fig:dg1dcorr}
\end{figure}

Other correlated observables of potential interest
are collected in Fig.~\ref{fig:dg1dcorr}. The top panel shows $g_1^{\Pqc}$ differential in
the transverse momentum $p_{T,\PQ\PaQ}$ of the HQ pair, the middle panel the correlation
in its azimuthal angle $2x\,dg_1^{\Pqc}/d\Delta\varphi$, i.e., $\Delta \varphi$ is the
azimuthal angle between the transverse momenta of the heavy quark and antiquark.
The quantity $2x\,dg_1^{\Pqc}/d\Delta R$, where $\Delta R\equiv \sqrt{(\Delta \varphi)^2 + (\Delta\eta)^2}$
denotes the HQ cone size variable, is illustrated in the bottom panel. Here,
$\Delta \eta$ refers to the difference in {\em pseudo}-rapidity of the charm-anticharm pair,
where $\eta=\frac{1}{2}\ln\left[(1+\cos(\theta))/(1-\cos(\theta))\right]$ with $\theta$ the angle of the HQ
relative to the $\Pggx\Pp$-axis in the c.m.s.\ frame. 
The histograms have been obtained with the MC code for the three sets of DSSV helicity densities
(best fit, min $\Delta \Pg$, and max $\Delta \Pg$) and the shaded bands correspond to the range allowed
by the DSSV uncertainty estimates. All results have been obtained at NLO accuracy 
for our standard choice of DIS kinematics, $x=0.001$ and $Q^2=\SI{10}{\GeV^2}$,
and setting the scale to $\mu^2= 4m^2+Q^2 + p_{T,\PQ\PaQ}^2$. In this case, we refrain from showing the
corresponding double-spin asymmetries as the distributions in Fig.~\ref{fig:dg1dcorr} are
presumably initially of more limited phenomenological interest at the EIC 
than the ones presented so far. We note that we have, as usual,
successfully reproduced all the respective unpolarized DIS results given in Ref.~\cite{Harris:1995tu}. 

All three distributions given in Fig.~\ref{fig:dg1dcorr} measure certain features of the
DIS HQ structure functions which are present for the first time beyond LO accuracy.
The distribution in the transverse momentum of the HQ pair (top panel) reveals the transverse momentum
of the additional light-parton jet which recoils against the HQ pair.
Since there is no such jet present at LO and in the virtual corrections at NLO accuracy, the
$p_{T,\PQ\PaQ}$-distribution peaks at small values, and the first bin is dominated by
``counter events'' in the MC, see also Ref.~\cite{Harris:1995tu}. Likewise, the distribution in
the azimuthal angle $\Delta \varphi$ between the produced charm quark and antiquark,
measured in the $\Pggx\Pp$-c.m.s.\ frame, peaks at the lowest order back-to-back configuration, i.e.,
for $\Delta\varphi=\pi$. The tails are produced by additional radiation of partons present
for the first time at NLO accuracy. Similar remarks apply to the cone size variable $\Delta R$
shown in the bottom panel, where again the LO distribution is a delta function at $\Delta R=\pi$.

Finally, we wish to comment on possible future additions to our suite of HQ electroproduction
codes. First and foremost, our MC code can be expanded in several directions. 
Following the unpolarized \texttt{HVQDIS} code \cite{Harris:1997zq}, the decay of
the HQs can be modeled by including hadronization, i.e., additional convolutions
with non-perturbative functions such as the charm-to-$\PD$-meson fragmentation function.
If necessary, one could go even one step further and include also the decay of the
heavy meson into experimentally observed electrons and muons; this was done,
for instance, in the case of HQ polarized hadroproduction \cite{Riedl:2009ye}. 
Another worthwhile addition would be to include also the lepton in the initial state in the MC
which would enable one to study all the observables discussed in this paper directly in
the laboratory rather than the $\Pggx\Pp$ c.m.s.\ frame; again, see the \texttt{HVQDIS} 
code \cite{Harris:1997zq} for an example.
Secondly, in close collaboration with experimentalists, one can certainly optimize the HQ DIS
observables presented in this paper by scanning for the best ranges in $x$, $Q^2$, $p_T$, etc.
Presumably this should wait until the the EIC is further along its way,
the machine parameters are confirmed, and experimental collaborations with final 
detector concepts are formed.

\section{Summary and Outlook \label{sec:summary}}
In this paper we have presented a new parton-level Monte Carlo
program at next-to-leading order accuracy in QCD
that allows one to study heavy flavor production 
in longitudinally polarized deep-inelastic scattering 
in terms of any infrared-safe differential distribution
for the structure function $g_1^{\PQ}$.
The full heavy flavor mass dependence is retained throughout all
calculations which makes this code particularly suited for
phenomenological studies at the future Electron-Ion Collider 
which can explore charm electroproduction 
at small-to-medium momentum fractions and 
photon virtualities not much larger than the charm quark mass.

We have validated the Monte Carlo generator against our 
results for fully inclusive heavy quark production
that where obtained previously with largely analytical methods.
Within the latter framework it is also possible to compare to
single-inclusive distributions differential in the transverse momentum 
or the rapidity of the observed heavy (anti)quark, albeit with
no experimental cuts. In addition, we have verified known unpolarized
results for heavy quark electroproduction that are available in the literature.

First phenomenological studies were carried out for various heavy quark 
distributions in the kinematic regime most
relevant for the future Electron-Ion Collider. 
Particular emphasis was devoted to the expected size of the
corresponding double-spin asymmetries and their sensitivity
to the still poorly constrained helicity gluon distribution.
Apart from single-inclusive distributions, we have also studied 
observables associated with the produced heavy quark pair 
including its invariant mass distribution and its correlation in azimuthal angle.
Theoretical uncertainties associated with the choice of the factorization scale
were studied using the important example of the single-inclusive 
transverse momentum distribution.
Here, it was found that the inclusion of next-to-leading order QCD 
corrections reduces the scale dependence significantly.

An additional benefit of now having a flexible Monte Carlo generator for
heavy quark polarized electroproduction at hand is the possibility to implement
and systematically explore relevant experimental cuts 
that might also help to enhance the
sensitivity to the helicity gluon density further. So far, apart
from the fully inclusive and transverse momentum differential
charm contributions to polarized deep-inelastic scattering, the
invariant mass spectrum appears to be the most promising observables
in this context.

At present, the code does not include any heavy quark decays but they 
can be straightforwardly implemented if needed, following already existing
frameworks and techniques in other codes. This can include even the decay of the
heavy mesons into experimentally observed electrons and muons.
The Monte Carlo generator can be also expanded in several other directions
most worthwhile might be the inclusion of the lepton beam kinematics
which would allow one to study heavy flavor electroproduction
not only in the virtual photon-hadron frame but in the
laboratory frame as well.

Finally, from a more theoretical perspective, 
the results presented in this and in our previous paper
are also relevant for setting up a 
general-mass variable flavor number scheme for helicity
dependent parton densities which might be more appropriate
whenever heavy flavors are studied in some asymptotic
region where the relevant scale, for instance, the
virtuality of the photon in deep-inelastic scattering,
is considerably larger than the heavy quark mass.

We plan to make the codes that were used in this and previous works publicly available.

\section*{Acknowledgments}
We are grateful to W.\ Vogelsang for helpful discussions and comments.
F. H. is supported by the European Research Council under the European
Unions Horizon 2020 research and innovation Programme (grant agreement no.\ 740006).
This study was supported in part by Deutsche Forschungsgemeinschaft (DFG) 
through the Research Unit FOR 2926 (project number 40824754). 

\begin{thebibliography}{0}%
\makeatletter
\providecommand \@ifxundefined [1]{%
 \@ifx{#1\undefined}
}%
\providecommand \@ifnum [1]{%
 \ifnum #1\expandafter \@firstoftwo
 \else \expandafter \@secondoftwo
 \fi
}%
\providecommand \@ifx [1]{%
 \ifx #1\expandafter \@firstoftwo
 \else \expandafter \@secondoftwo
 \fi
}%
\providecommand \natexlab [1]{#1}%
\providecommand \enquote  [1]{``#1''}%
\providecommand \bibnamefont  [1]{#1}%
\providecommand \bibfnamefont [1]{#1}%
\providecommand \citenamefont [1]{#1}%
\providecommand \href@noop [0]{\@secondoftwo}%
\providecommand \href [0]{\begingroup \@sanitize@url \@href}%
\providecommand \@href[1]{\@@startlink{#1}\@@href}%
\providecommand \@@href[1]{\endgroup#1\@@endlink}%
\providecommand \@sanitize@url [0]{\catcode `\\12\catcode `\$12\catcode
  `\&12\catcode `\#12\catcode `\^12\catcode `\_12\catcode `\%12\relax}%
\providecommand \@@startlink[1]{}%
\providecommand \@@endlink[0]{}%
\providecommand \url  [0]{\begingroup\@sanitize@url \@url }%
\providecommand \@url [1]{\endgroup\@href {#1}{\urlprefix }}%
\providecommand \urlprefix  [0]{URL }%
\providecommand \Eprint [0]{\href }%
\providecommand \doibase [0]{http://dx.doi.org/}%
\providecommand \selectlanguage [0]{\@gobble}%
\providecommand \bibinfo  [0]{\@secondoftwo}%
\providecommand \bibfield  [0]{\@secondoftwo}%
\providecommand \translation [1]{[#1]}%
\providecommand \BibitemOpen [0]{}%
\providecommand \bibitemStop [0]{}%
\providecommand \bibitemNoStop [0]{.\EOS\space}%
\providecommand \EOS [0]{\spacefactor3000\relax}%
\providecommand \BibitemShut  [1]{\csname bibitem#1\endcsname}%
\let\auto@bib@innerbib\@empty
%</preamble>
\end{thebibliography}%


\begin{thebibliography}{99}
%
%
%
%
%
\bibitem{Adolph:2012ca} 
  C.~Adolph {\it et al.} [COMPASS Collaboration],
  %
  Phys.\ Rev.\ D {\bf 87}, 052018 (2013).
  %
%
%
\bibitem{Nocera:2013yia} 
  E.~R.~Nocera,
  %
  PoS DIS {\bf 2013}, 211 (2013).
  %
%
\bibitem{Boer:2011fh}
  D.~Boer {\it et al.},
  %
  {\tt arXiv:1108.1713};
%
  A.~Accardi {\it et al.},
  %
  Eur.\ Phys.\ J.\ A {\bf 52}, 268 (2016);
  %
%
  E.~C.~Aschenauer {\it et al.},
  %
  %
  Rept.\ Prog.\ Phys.\ \textbf{82}, 024301 (2019);
%
  R.~Abdul Khalek {\it et al.}, 
  {\tt arXiv:2103.05419}.
  %
%
\bibitem{Aidala:2012mv} 
  For a review, see, e.g., C.~A.~Aidala, S.~D.~Bass, D.~Hasch, and G.~K.~Mallot,
  %
  Rev.\ Mod.\ Phys.\  {\bf 85}, 655 (2013).
  %
%
\bibitem{deFlorian:2008mr} 
  D.~de Florian, R.~Sassot, M.~Stratmann, and W.~Vogelsang,
  %
  Phys.\ Rev.\ Lett.\  {\bf 101}, 072001 (2008);
  %
  %
  Phys.\ Rev.\ D {\bf 80}, 034030 (2009).
  %
%
\bibitem{deFlorian:2014yva} 
  D.~de Florian, R.~Sassot, M.~Stratmann, and W.~Vogelsang,
  %
  Phys.\ Rev.\ Lett.\  {\bf 113}, 012001 (2014).
  %
%
\bibitem{Blumlein:2010rn} 
  J.~Blumlein and H.~Bottcher,
  %
  Nucl.\ Phys.\ B {\bf 841}, 205 (2010);
%
  E.~Leader, A.~V.~Sidorov, and D.~B.~Stamenov,
  %
  Phys.\ Rev.\ D {\bf 82}, 114018 (2010);
  %
%
  E.~R.~Nocera {\it et al.} [NNPDF Collaboration],
  %
  Nucl.\ Phys.\ B {\bf 887}, 276 (2014);
  %
%
  N.~Sato {\it et al.} [Jefferson Lab Angular Momentum Collaboration],
  %
  Phys.\ Rev.\ D {\bf 93}, 074005 (2016);
  %
  D.~De Florian, G.~A.~Lucero, R.~Sassot, M.~Stratmann, and W.~Vogelsang,
  %
  Phys.\ Rev.\ D {\bf 100}, 114027 (2019).
  %
%
\bibitem{Aschenauer:2012ve} 
  Some exploratory studies based on mock EIC data but using only massless quarks
  can be found in: 
  E.~C.~Aschenauer, R.~Sassot, and M.~Stratmann,
  %
  Phys.\ Rev.\ D {\bf 86}, 054020 (2012);
  %
  %
  Phys.\ Rev.\ D {\bf 92}, 094030 (2015);
  %
  R.~D.~Ball {\it et al.} [NNPDF Collaboration],
  %
  Phys.\ Lett.\ B {\bf 728}, 524 (2014);
  %
  E.~C.~Aschenauer, I.~Borsa, R.~Sassot, and C.~Van Hulse,
  %
  Phys.\ Rev.\ D {\bf 99}, 094004 (2019);
  %
  %
  I.~Borsa, G.~Lucero, R.~Sassot, E.~C.~Aschenauer, and A.~S.~Nunes,
  %
  Phys.\ Rev.\ D \textbf{102}, 094018 (2020);
  %
  Y.~Zhou {\it et al.}, 
  %
  {\tt arXiv:2105.04434}.
%
\bibitem{H1:2018flt} 
  H.~Abramowicz {\it et al.} [H1 and ZEUS Collaborations],
  %
  Eur.\ Phys.\ J.\ C {\bf 73}, 2311 (2013);
  %
  Eur.\ Phys.\ J.\ C {\bf 78}, 473 (2018).
  %
%
\bibitem{Vogelsang:1990ug} 
  See, e.g., W.~Vogelsang,
  %
  Z.\ Phys.\ C {\bf 50}, 275 (1991);
  %
  %
  S.~Moch, J.~A.~M.~Vermaseren, and A.~Vogt,
  %
  Phys.\ Lett.\ B \textbf{748}, 432-438 (2015).
%
\bibitem{Bojak:2001fx}
 I.~Bojak and M.~Stratmann,
  %
  Phys.\ Rev.\ D {\bf 67}, 034010 (2003).
  %
%
\bibitem{Riedl:2009ye} 
  J.~Riedl, A.~Schafer, and M.~Stratmann,
  %
  Phys.\ Rev.\ D {\bf 80}, 114020 (2009).
  %
%
\bibitem{Bojak:1998bd} 
  I.~Bojak and M.~Stratmann,
  %
  Phys.\ Lett.\ B {\bf 433}, 411 (1998);
  %
  %
  Nucl.\ Phys.\ B {\bf 540}, 345 (1999),
  Erratum: [Nucl.\ Phys.\ B {\bf 569}, 694 (2000)];
  %
  Z.~Merebashvili, A.~P.~Contogouris, and G.~Grispos,
  %
  Phys.\ Rev.\ D {\bf 62}, 114509 (2000),
  Erratum: [Phys.\ Rev.\ D {\bf 69}, 019901 (2004)].
  %
%
\bibitem{Riedl:2012qc} 
  J.~Riedl, M.~Stratmann, and A.~Schafer,
  %
  Eur.\ Phys.\ J.\ C {\bf 73}, 2360 (2013).
  %
%
\bibitem{Hekhorn:2018ywm} 
  F.~Hekhorn and M.~Stratmann,
  %
  Phys.\ Rev.\ D {\bf 98}, 014018 (2018).
  %
%
\bibitem{Watson:1981ce} 
  A.~D.~Watson,
  %
  Z.\ Phys.\ C {\bf 12}, 123 (1982);
  %
%
  M.~Gluck, E.~Reya, and W.~Vogelsang,
  %
  Nucl.\ Phys.\ B {\bf 351}, 579 (1991).
  %
%
\bibitem{Smith:1991pw} 
%
  R.~K.~Ellis and P.~Nason,
  %
  Nucl.\ Phys.\ B {\bf 312}, 551 (1989);
  %
  J.~Smith and W.~L.~van Neerven,
  %
  Nucl.\ Phys.\ B {\bf 374}, 36 (1992).
  %
%
\bibitem{Laenen:1992zk} 
  E.~Laenen, S.~Riemersma, J.~Smith, and W.~L.~van Neerven,
  %
  Nucl.\ Phys.\ B {\bf 392}, 162 (1993).
  %
%
\bibitem{Kauer:2001sp} 
  G.~P.~Lepage,
  %
  J.\ Comput.\ Phys.\  {\bf 27}, 192 (1978);
  %
  N.~Kauer and D.~Zeppenfeld,
  %
  Phys.\ Rev.\ D {\bf 65}, 014021 (2002); the {\sc DVegas} code can
  be obtained from {\tt https://dvegas.hepforge.org}.
  %
%
\bibitem{tHooft:1972tcz} 
  G.~'t Hooft and M.~J.~G.~Veltman,
  %
  Nucl.\ Phys.\ B {\bf 44}, 189 (1972);
  %
%
  P.~Breitenlohner and D.~Maison,
  %
  Commun.\ Math.\ Phys.\  {\bf 52}, 11 (1977); ibid. 39 (1977);
  ibid.\ 55 (1977).
  %
%
\bibitem{Ellis:1980wv} 
  R.~K.~Ellis, D.~A.~Ross, and A.~E.~Terrano,
  %
  Nucl.\ Phys.\ B {\bf 178}, 421 (1981).
  %
%
\bibitem{Mangano:1991jk} 
  M.~L.~Mangano, P.~Nason, and G.~Ridolfi,
  %
  Nucl.\ Phys.\ B {\bf 373}, 295 (1992).
  %
%
\bibitem{Frixione:1993dg} 
  S.~Frixione, M.~L.~Mangano, P.~Nason, and G.~Ridolfi,
  %
  Nucl.\ Phys.\ B {\bf 412}, 225 (1994).
  %
%
\bibitem{Harris:1995tu} 
  B.~W.~Harris and J.~Smith,
  %
  Nucl.\ Phys.\ B {\bf 452}, 109 (1995).
  %
%
\bibitem{Laenen:1992xs} 
  E.~Laenen, S.~Riemersma, J.~Smith, and W.~L.~van Neerven,
  %
  Nucl.\ Phys.\ B {\bf 392}, 229 (1993).
  %
%
\bibitem{Riemersma:1994hv} 
  S.~Riemersma, J.~Smith, and W.~L.~van Neerven,
  %
  Phys.\ Lett.\ B {\bf 347}, 143 (1995).
  %
%
\bibitem{Harris:1995pr} 
  B.~W.~Harris and J.~Smith,
  %
  Phys.\ Lett.\ B {\bf 353}, 535 (1995),
  Erratum: [Phys.\ Lett.\ B {\bf 359}, 423 (1995)].
  %
%
\bibitem{Harris:1997zq} 
  B.~W.~Harris and J.~Smith,
  %
  Phys.\ Rev.\ D {\bf 57}, 2806 (1998).
  %
%
\bibitem{Laenen:1998kp} 
  E.~Laenen and S.~O.~Moch,
  %
  Phys.\ Rev.\ D {\bf 59}, 034027 (1999);
  %
%
  T.~O.~Eynck and S.~O.~Moch,
  %
  Phys.\ Lett.\ B {\bf 495}, 87 (2000);
  %
%
  H.~Kawamura, N.~A.~Lo Presti, S.~Moch, and A.~Vogt,
  %
  Nucl.\ Phys.\ B {\bf 864}, 399 (2012).
  %
%
%
\bibitem{Brodsky:1980pb} 
  S.~J.~Brodsky, P.~Hoyer, C.~Peterson, and N.~Sakai,
  %
  Phys.\ Lett.\  {\bf 93B}, 451 (1980);
%
  %
%
%
  S.~J.~Brodsky, C.~Peterson, and N.~Sakai,
  %
  Phys.\ Rev.\ D {\bf 23}, 2745 (1981).
%
  %
%
\bibitem{Buza:1996wv} 
  M.~Buza, Y.~Matiounine, J.~Smith, and W.~L.~van Neerven,
  %
  Eur.\ Phys.\ J.\ C {\bf 1}, 301 (1998);
  %
%
  A.~Chuvakin, J.~Smith, and W.~L.~van Neerven,
  %
  Phys.\ Rev.\ D {\bf 61}, 096004 (2000);
  %
%
  J.~Blumlein, A.~De Freitas, C.~Schneider, and K.~Schonwald,
  %
  Phys.\ Lett.\ B {\bf 782}, 362 (2018);
  %
%
%
  S.~Alekhin, J.~Bl\"umlein, and S.~Moch,
%
  Phys.\ Rev.\ D \textbf{102}, 054014 (2020).
%
\bibitem{Butterworth:2015oua} 
  For a review of the PDF4LHC recommendations, see, J.~Butterworth {\it et al.},
  %
  J.\ Phys.\ G {\bf 43}, 023001 (2016);
  %
  see also, 
%
  S.~Forte, E.~Laenen, P.~Nason, and J.~Rojo,
  %
  Nucl.\ Phys.\ B {\bf 834}, 116 (2010) and references therein.
  %
%
\bibitem{Accardi:2016ndt} 
  For a critical appraisal, see, A.~Accardi {\it et al.},
  %
  Eur.\ Phys.\ J.\ C {\bf 76}, 471 (2016).
  %
%
\bibitem{Buza:1996xr} 
  M.~Buza, Y.~Matiounine, J.~Smith, R.~Migneron, and W.~L.~van Neerven,
  %
  Nucl.\ Phys.\ B {\bf 472}, 611 (1996);
  %
  M.~Buza, Y.~Matiounine, J.~Smith, and W.~L.~van Neerven,
  %
  Nucl.\ Phys.\ B {\bf 485}, 420 (1997);
  %
 %
 %
  J.~Bl\"umlein, A.~De Freitas, M.~Saragnese, C.~Schneider, and K.~Sch\"onwald,
%
  {\tt arXiv:2105.09572}.
%
%
%
%
\bibitem{Riedl:2014ywt} 
  J.~Riedl, Ph.D.~Thesis, University of Regensburg, 2014.
  %
%
\bibitem{Phaf:2001gc} 
  A general formulation of the dipole subtraction method for
  massive partons in QCD has been developed in:    
  L.~Phaf and S.~Weinzierl,
  %
  JHEP {\bf 0104}, 006 (2001);
  %
  S.~Catani, S.~Dittmaier, M.~H.~Seymour, and Z.~Trocsanyi,
  %
  Nucl.\ Phys.\ B {\bf 627}, 189 (2002).
  %
%
\bibitem{ref:phd} 
%
  F.~Hekhorn, Ph.D.~Thesis, University of T\"ubingen, Aug.\ 2019,
  %
%
  \texttt{arXiv:1910.01536}.
  %
%
\bibitem{Vogelsang:1995vh} 
  W.~Vogelsang,
  %
  Phys.\ Rev.\ D {\bf 54}, 2023 (1996);
  %
%
  %
  Nucl.\ Phys.\ B {\bf 475}, 47 (1996);
  %
%
  R.~Mertig and W.~L.~van Neerven,
  %
  Z.\ Phys.\ C {\bf 70}, 637 (1996).
  %
%
%
%
\bibitem{Martin:2009iq} 
  A.~D.~Martin, W.~J.~Stirling, R.~S.~Thorne, and G.~Watt,
  %
  Eur.\ Phys.\ J.\ C {\bf 63}, 189 (2009).
  %
%
\bibitem{Cacciari:1993mq}
  M.~Cacciari and M.~Greco,
  %
  Nucl.\ Phys.\ B \textbf{421}, 530 (1994);
  %
  M.~Cacciari, M.~Greco, and P.~Nason,
  %
  JHEP \textbf{05}, 007 (1998);
  %
  M.~Cacciari, S.~Frixione, and P.~Nason,
  %
  JHEP \textbf{03}, 006 (2001).
%
\end{thebibliography}
\end{document}